\documentclass{aa}

\usepackage[caption=false]{subfig}
\usepackage{float}
\usepackage{graphicx}	
\usepackage{amsmath}	
\usepackage{amssymb}
\usepackage{xcolor}
\usepackage{booktabs}
\usepackage{enumerate}
\usepackage{dirtytalk}
\usepackage{float}
\usepackage{enumitem}
\usepackage{booktabs}
\usepackage{lipsum} 
\usepackage{multirow}
\usepackage{makecell}
\usepackage[english]{babel}
\usepackage{comment}
\usepackage{multirow}
\usepackage{footnote}
\usepackage[normal]{threeparttable}
\usepackage{hyperref}

\defcitealias{Benneke2024}{B24}
\defcitealias{Holmberg2024}{HM24}

\begin{document}

\title{The atmospheric composition of TOI-270~d}

   \author{Savvas Constantinou
          \inst{1}\fnmsep
          \and
          Nikku Madhusudhan\inst{1}
          \and 
          M\aa ns Holmberg \inst{1, 2}
          }

   \institute{\inst{1} Institute of Astronomy, University of Cambridge, Madingley Road, Cambridge CB3 0HA, UK\\
              \inst{2} Space Telescope Science Institute, 3700 San Martin Drive, Baltimore, MD 21218, USA \\
            \email{sc938@cam.ac.uk} \\
            \email{nmadhu@ast.cam.ac.uk}
             }

\date{}

\abstract{The first explorations of temperate sub-Neptune exoplanets have been the hallmark of early JWST observations. The bulk properties of such planets are consistent with a range of possible internal structures, which can be distinguished through their interactions with the observable atmospheres. JWST observations of TOI-270~d, a temperate sub-Neptune, have previously led to contrasting conclusions: either a Hycean world, possessing a liquid water ocean, or a mixed-envelope sub-Neptune, where high temperatures prevent a liquid ocean and lead to a high mean molecular weight atmosphere. In order to resolve this uncertainty, we present a comprehensive retrieval analysis of TOI-270~d using recent NIRISS and NIRSpec transit spectroscopy across $\sim$1-5 $\mu$m. We find that prior inferences of a mixed envelope were affected by specific modelling choices leading to a high terminator temperature and high mean-molecular weight in the atmosphere. We confirm an H$_2$-rich atmosphere in TOI-270~d and present revised constraints on the molecular log-mixing ratios and maximal detection significances of CH$_4$ at $-1.86^{+0.30}_{-0.29}$ (6.4 $\sigma$), CO$_2$ at $-1.71^{+0.38}_{-0.66}$ (3.9 $\sigma$), H$_2$O at $-1.88^{+0.78}_{-4.13}$ (2.1 $\sigma$) and CS$_2$ at $-4.74^{+0.65}_{-1.10}$ (2.0 $\sigma$), with a terminator temperature of $323^{+58}_{-52}$~K at 10~mbar. We also find tentative evidence for more complex methyl-bearing species such as C$_2$H$_6$ and/or DMS at a 2.1-2.5 $\sigma$ level. The present constraints are consistent with TOI-270~d being a Hycean or dark Hycean world, with planet-wide or nightside liquid water oceans. However, more observations are required to verify the present findings and robustly constrain the atmospheric conditions and internal structure of TOI-270~d.}


   \maketitle

\section{Introduction} \label{sec:introduction}

The nature of sub-Neptune planets ranks among the biggest open questions in the study of extrasolar planets. Based on the current population of detected exoplanets, sub-Neptunes are likely the most common type of planet in the solar neighbourhood \citep{fulton2018, Cloutier2020}. Despite their prevalence, however, their actual nature remains a mystery. With radii between those of Neptune and Earth, such planets have no solar system equivalent. Moreover, their bulk properties are consistent with several internal structures, including those based on solar system archetypes, like mini-Neptunes and super-Earths.

For temperate sub-Neptunes, nestled in the parameter space of plausible internal structures are configurations with habitable surface conditions \citep[e.g.][]{Madhusudhan2020, Rigby2024}. One such configuration pertains to Hycean planets \citep{Madhusudhan2021}, which consist of H$_2$-rich atmospheres overlying deep liquid H$_2$O oceans. As such, in addition to posing exciting questions in planetary science, temperate sub-Neptunes present a unique opportunity to accelerate the search for life beyond Earth. 

Unlocking the immense discovery space in the sub-Neptune regime requires atmospheric characterisation. The atmospheres of sub-Neptune are expected to be significantly affected by their internal structures. In particular, the presence, nature and depth of a planetary surface can alter the interplay of thermochemical, photochemical and transport processes shaping the atmosphere \citep{Yu2021, Hu2021, Tsai2021, Madhusudhan2023a}, resulting in drastically different atmospheric compositions. This means that with atmospheric observations of sufficient quality, it would be possible to break the degeneracies in internal structures.

Despite their immense promise, probing the atmospheres of temperate sub-Neptunes constitutes a significant challenge, as their relatively small sizes and corresponding planet-star size ratios lead to small spectroscopic signals. This challenge was finally overcome with the arrival of the James Webb Space Telescope (JWST), with a sensitivity high enough to capture the small spectroscopic signals and a wavelength coverage large enough to enable robust characterisation. Sub-Neptune planets have been an early focus of JWST, with several such planets under observation as part of the first few cycles.

The early focus of JWST on sub-Neptunes has already led to important breakthroughs. The first temperate sub-Neptune observed was K2-18~b \citep{Montet2015, Sarkis2018}, a habitable zone sub-Neptune with a mass of $8.63 \pm 1.35$~M$_\oplus$, radius of $2.61 \pm 0.09$~M$_\oplus$, and an equilibrium temperature of $\sim$250-300~K, for Bond albedo values between 0 and 0.5 \citep{Cloutier2019, Benneke2019}. Observed with both NIRISS and NIRSpec instruments, these observations led to the first detections of CH$_4$, and CO$_2$ in a temperate exoplanetary atmosphere \citep{Madhusudhan2023b}.  

Following the observations of K2-18~b, JWST was used to observe TOI-270~d \citep{Gunther2019}, another temperate sub-Neptune. TOI-270~d has a mass almost half that of K2-18~b, $M_\mathrm{p} = 4.78 \pm 0.43 M_\oplus$, a slightly smaller radius of $2.133 \pm 0.058 R_\oplus$ and a higher equilibrium temperature between $\sim$330-390~K for albedos of 0-0.5 \citep{van2021masses}. Transmission spectroscopy with JWST once again led to confident detections of CH$_4$ and CO$_2$, with the additional detection of CS$_2$ and potentially H$_2$O \citep{Holmberg2024, Benneke2024}.

Beyond novel detections, JWST transmission spectra can shed light on the internal structures of planets like K2-18~b and TOI-270~d. A planet's interior, whether it is a deep atmosphere for mini-Neptunes, the deep ocean of a Hycean world or a rocky surface of a super-Earth, is expected to interact with the observable atmosphere. Such interactions include recycling of thermochemical products in deeper layers in mini-Neptunes, outgassing of volcanic gases on super-Earths, and dissolution of specific species in liquid water oceans of Hycean worlds \citep{Yu2021, Hu2021, Tsai2021, Madhusudhan2023a}. In the case of K2-18~b, its atmospheric properties were found to be consistent with a Hycean planet scenario, with significant CH$_4$, CO$_2$, and a dearth of NH$_3$ due to its dissolution in an H$_2$O ocean \citep{Madhusudhan2023b}.

The Hycean scenario was also invoked for TOI-270~d by \citet[][]{Holmberg2024}, hereafter \citetalias{Holmberg2024}, using HST WFC3 and JWST NIRSpec observations in the 1.1-1.7~$\mu$m and 2.7-5.2~$\mu$m ranges. Specifically, \citetalias{Holmberg2024} obtained atmospheric abundance constraints for CH$_4$ and CO$_2$ of $\sim$0.1-1~\% while also finding a significant relative depletion of NH$_3$, comparable to those of K2-18~b \citep{Madhusudhan2023b}. Such abundances, along with a terminator temperature estimate of $\sim$$300 \pm 70$~K, are consistent with TOI-270~d being a Dark Hycean planet, characterised by a liquid ocean on its cooler nightside.

However, the same JWST NIRSpec observations of TOI-270~d, paired with additional NIRISS observations to reach a combined 1-5~$\mu$m wavelength coverage, have been interpreted by \citet{Benneke2024}, hereafter \citetalias{Benneke2024}, to instead indicate that TOI-270~d is a mixed-envelope sub-Neptune. A mixed-envelope scenario for warm sub-Neptunes was explored by \citet{Nixon2021}, which involves higher temperatures preventing a liquid water ocean, with the H$_2$O instead being supercritical and well-mixed with the H$_2$-rich atmosphere \citep{Nixon2021}. \citetalias{Benneke2024} suggested that this scenario would result in a higher mean molecular weight (MMW) in the atmosphere compared to the Hycean scenario. Such an atmosphere can still produce readily observable spectral features comparable to the Hycean scenario, as the higher temperature offsets the higher MMW. 

The source of the discrepancy between the two different interpretations of \citetalias{Benneke2024} and \citetalias{Holmberg2024} are differences in the retrieved atmospheric temperature and mean molecular weight (MMW). The temperatures are constrained directly by the respective retrieval frameworks, with \citetalias{Holmberg2024} finding a temperature of $289^{+80}_{-75}$~K at 10~mbar when considering one offset and a dual transit model, the case analogous to that of \citetalias{Benneke2024}. By contrast, \citetalias{Benneke2024} instead find a significantly higher temperature of $385^{+44}_{-42}$~K with their free chemistry retrievals. Atmospheric MMW constraints are indirectly inferred from the abundance constraints obtained from individual molecular species. \citetalias{Benneke2024} infer an MMW constraint of $5.47^{+1.25}_{-1.14}$~amu, driven by high mixing ratio constraints for CH$_4$ and CO$_2$ which are confidently detected, and H$_2$O which is tentatively detected at $2.6 \sigma$. \citetalias{Holmberg2024} obtain mixing ratio constraints that are lower than those obtained by \citetalias{Benneke2024} and correspond to a MMW of $\sim$2.4-4.8~amu. The source of this discrepancy was not known, as the two works used different sets of observations, different reduction pipelines for the observations and different atmospheric models for the retrievals.

The present work has three main objectives. The first is to resolve the discrepancy between the above interpretations of the transmission spectrum of TOI-270~d. Second, we combine a new reduction of the available NIRISS observations with the existing NIRSpec reduction of \citetalias{Holmberg2024} to present revised constraints on atmospheric properties. Third, using this combined dataset we provide new insights into the atmospheric composition of TOI-270~d, including the potential presence of methyl-bearing species.

We begin by discussing the observations used in this work in Section \ref{sec:observations}, including the reduction process for the NIRISS observations. We then discuss the atmospheric retrieval methodology followed in this work in Section \ref{sec:methods}, including both the models used to reproduce prior findings as well as the canonical model of the present work. We begin our retrieval analysis by reproducing findings in literature in Section \ref{sec:benneke_reproduction}, identifying the aspects of the retrieved atmospheric model which can affect atmospheric constraints. We present the results obtained with our canonical atmospheric retrieval model in Section \ref{sec:canonical_retrievals}. We lastly summarise the present work and discuss the implications in Section \ref{sec:discussion}.

\section{Observations} \label{sec:observations}
In the present work, we consider transit observations of TOI-270~d obtained with JWST NIRISS SOSS and NIRSpec G395H. Together, these observations achieve a near-continuous coverage across a wavelength range of $0.6-5.2~\mu$m. The observations were originally envisaged in multiple JWST programs, GO 3557 (PI: N. Madhusudhan) and GO 4098 (PI: B. Benneke), and were implemented as part of the latter.

We present a new analysis of the NIRISS observation, obtained on February 6, 2024, using the SUBSTRIP256 subarray and the NISRAPID readout pattern. The observation consists of 868 integrations, with three groups per integration, spanning a total of 5.3 hours. In addition, we also carry out retrieval analysis of data presented in the literature. Specifically, we consider the combined NIRISS and NIRSpec observations presented by \citetalias{Benneke2024} in Section~\ref{sec:benneke_reproduction}. For the rest of the work, including our canonical retrievals, we combine the NIRISS spectrum presented below with the NIRSpec data presented by \citetalias{Holmberg2024}.

\subsection{NIRISS SOSS data reduction}\label{subsec:niriss_reduction}

For Stages 1 and 2, we used the JWST Science Calibration Pipeline \citep{Bushouse2020} in combination with a custom 1/f noise correction step similar to \cite{Madhusudhan2023b}. In Stage 3, we used a variant of JExoRES \citep{Holmberg2023} to extract the spectrum. First, we apply the standard Stage~1 steps: data quality initialisation, saturation checking, and superbias subtraction. We do not perform the jump step, due to few groups, or the dark current subtraction step, as this can induce additional noise \citep{Radica2023}. Instead, outliers due to cosmic ray hits are flagged in Stage 3. We note that the spectrum did not change significantly with the dark current subtraction step turned on or off. For the 1/f noise correction, we first build a linear model of the detector counts at the group level,
\begin{equation}
    M_{ij}^{(k)}(t) = \left(m_{ij}^{(k)} - b_{ij}^{(k)}\right) T(t) + b_{ij}^{(k)} + c_{ij}\,,
\end{equation}
where $i, j$ are the pixel coordinates, $k$ is the group number, $m_{ij}^{(k)}$ is the median count of the $k$th group (across all out-of-transit integrations), $b_{ij}^{(k)}$ is the background count of the $k$th group, $T(t)$ is the normalised white light curve, and $c_{ij}$ is the bias count. Note that both the median and background counts are bias subtracted. We estimated the bias of each pixel by the constant obtained from fitting a linear polynomial to the median counts (ramp fitting). To avoid non-linear detector effects, we corrected the bias values within the traces of the two spectral orders ($\pm$ 20 pixels) by replacing these regions with interpolated values. For the present observation, we find bias values of around 22~DN (after performing the superbias subtraction step). The background is modelled as described below, using the standard STScI model background, but at the group level. 

Once we constructed the above model, we temporarily subtracted it from the data and used the residuals to reveal the 1/f noise, which is then subtracted from the data. For this, we used the mean of the residuals per detector column and integration while masking the traces of the two spectral orders ($\pm$ 20 pixels), field star contamination, and outliers (via sigma clipping). We did this correction for even and odd rows separately \citep{Radica2023}. Note that, initially, the white light curve is not known at Stage 1, which is why we iterated all steps of the data reduction, starting with a constant white light curve, filled with ones. After Stage 3, we obtained the actual white light curve, which we used in the next iteration. We iterated Stages 1-3 twice after the initial pass (without 1/f noise correction), at which point the white light curve did not change significantly. We note that the present 1/f noise correction is different to the methods used in \cite{Benneke2024}. Finally, we finished Stages 1 and 2 by performing the non-linearity correction, ramp fitting, and flat field correction. 

\begin{figure}
\noindent
\includegraphics[angle=0,width=\columnwidth]{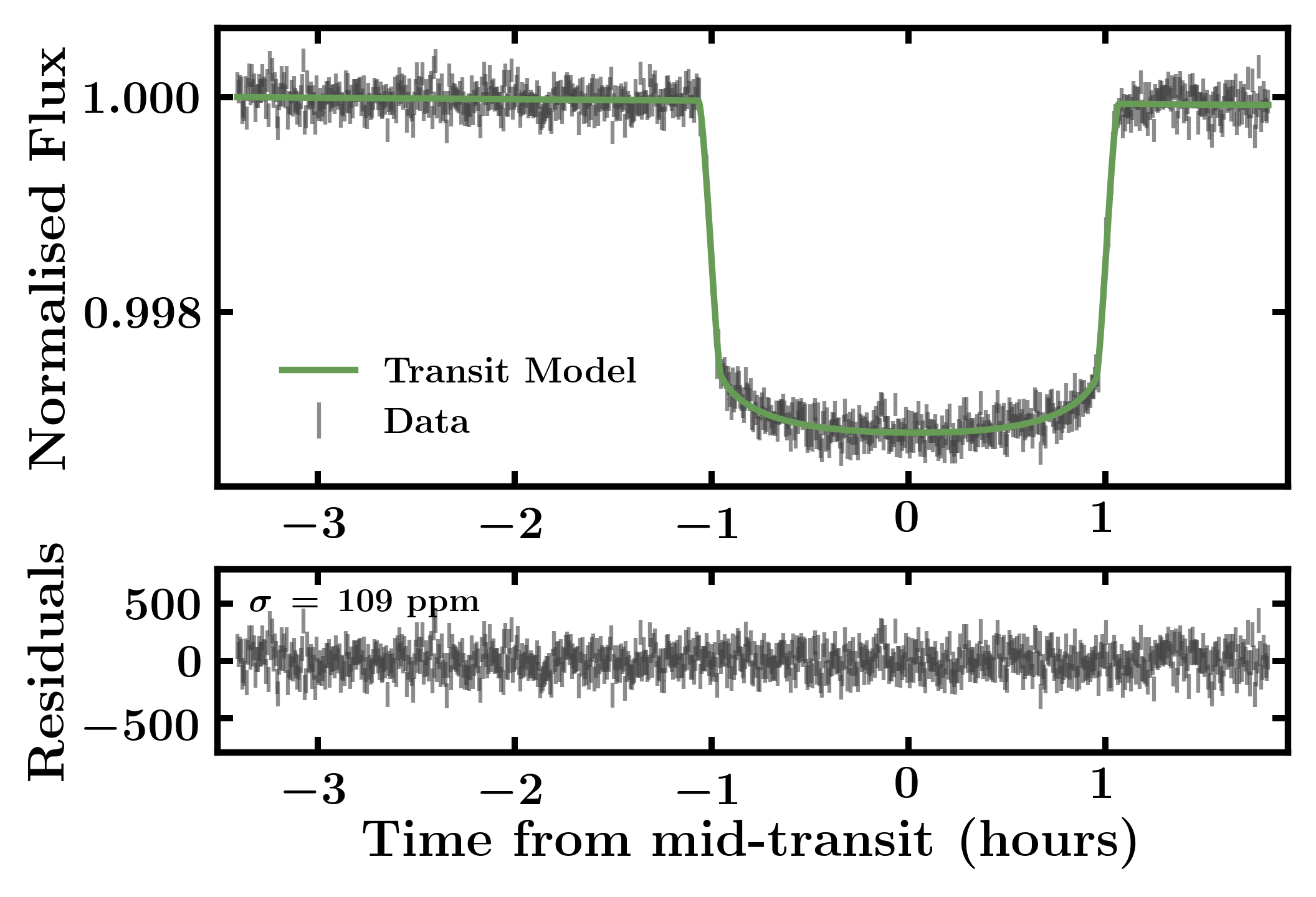}
\caption{White light curve for the transit of TOI-270~d observed with NIRISS SOSS, order 1. The data and best-fit model are shown in black and green, respectively. The standard deviation of the residuals is 109 ppm, corresponding to 2.3 times the expected noise level, similar to other NIRISS SOSS observations. The parameter constraints from the model fit are provided in Table~\ref{tab:wlc_params}.}
\label{fig:wlc}
\end{figure}

For Stage 3, we first searched for outliers in the time series of each pixel. To do this, we subtracted a median filtered version of the pixel time series (with a window size of 7 integrations) and performed sigma clipping on the residuals. For each integration, we applied a binary dilation to the outlier mask to reject neighbouring pixels. We replaced these outliers by linearly interpolating in the spectral direction. Next, we obtained the wavelength calibration and an initial order tracing using PASTASOSS \citep{Baines2023a, Baines2023b}. We then refined the order tracing according to the method presented by \cite{Holmberg2023}. For the background subtraction, we scaled and subtracted the standard STScI model background using the median of all integrations in a small rectangular region ($ x \in [250, 500], \, y \in [210, 250]$). To extract the spectra, we used a box extraction with an aperture of 36 pixels, as the contamination between the two orders is insignificant \citep{Darveau-Bernier2022, Radica2023, Holmberg2023}. As a final quality control, we searched and removed outliers from the extracted time series spectrum, as described above, with a threshold set to $4\sigma$. Any outliers found at this stage were removed from further analysis. Finally, we summed up the flux from all wavelength channels of order~1 to make the white light curve, shown in Figure~\ref{fig:wlc}.

We fit the light curves with the batman transit model \citep{Kreidberg2015} using MultiNest \citep{Buchner2014}, assuming a circular orbit and a fixed orbital period of 11.38014 days \citep{Mikal-Evans2023}. We adopted the two-parameter quadratic limb-darkening law using the parameterisation and priors by \cite{Kipping2013}. We also fit for an uncertainty inflation parameter and a linear baseline trend (using two additional parameters). First, we performed the white light curve fitting with normal priors for the scaled semi-major axis $a/R_*$ and orbital inclination $i$ from the dual-transit case in \cite{Holmberg2024}, $a/R_* = 42.34 \pm 0.30$ and $i = 89.748 \pm 0.054$. We used uninformative uniform priors for other parameters. The resulting transit parameters are provided in Table~\ref{tab:wlc_params}. Next, we binned the spectroscopic light curves to $R \approx 50$ and performed the transit model fitting of these to obtain the spectrum. We did this separately for each spectral order while fixing the orbital parameters to the values obtained from the white light curve fitting. For visualisation, we also generated spectra with $R \approx 12$ and $R \approx 25$.

Due to the presence of correlated noise, typical of NIRISS SOSS observations \citep{Holmberg2023}, we estimated an empirical covariance matrix to be used in the likelihood of the retrieval analysis. We do this via
\begin{equation}
    \Sigma_{ij} = \frac{\sigma_i \sigma_j}{\tilde \sigma_i \tilde \sigma_j} \tilde \Sigma_{ij}\,,
\end{equation}
where $\tilde \Sigma_{ij}$ represent the empirical covariance of the light curves between the $i$th and $j$th spectral channels, and where $\sigma_i$ and $\tilde \sigma_i$ are the transit depth uncertainty (from the light curve fitting) and light curve scatter of the $i$th spectral channel, respectively. Note that the coefficients $(\sigma_i \sigma_j) / (\tilde \sigma_i \tilde \sigma_j)$ all agree within a few per cent when using normalised light curves, as this approximates the constant $(J^T J)^{-1}_{\delta \delta}$, where $J$ is the gradient of the transit model \citep{Holmberg2023}. 

\begin{figure}
\noindent
\includegraphics[angle=0,width=0.97\columnwidth]{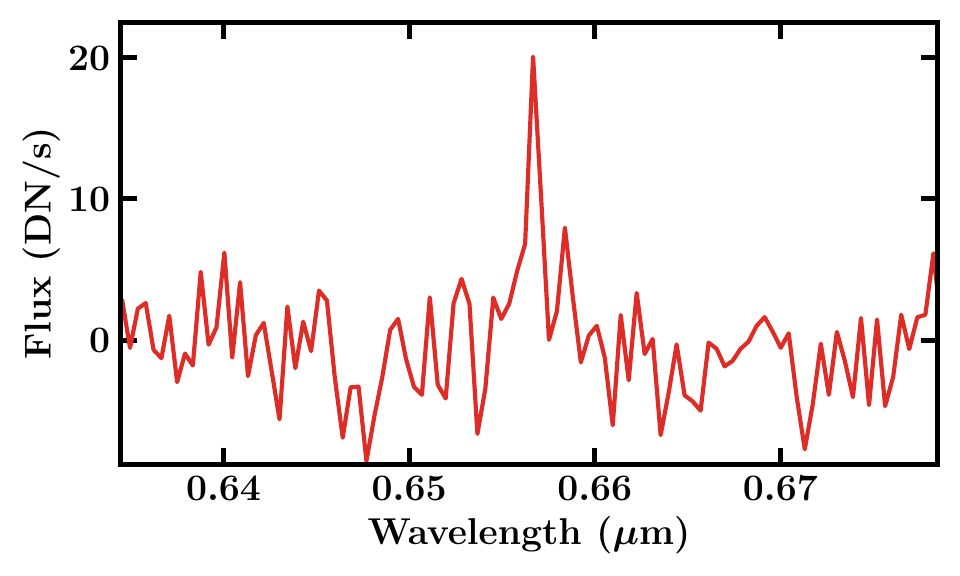}
\vspace{-1mm}
\caption{Excess H$\alpha$ emission from a minor flare. Note that the peak emission aligns with the small flux increase seen in the white light curve just before mid transit. Ultimately, this does not significantly impact the final transmission spectrum, as discussed in Section~\ref{subsec:niriss_reduction}.}
\label{fig:flare}
\end{figure}

Finally, it is worth noting that \citetalias{Benneke2024} found no flare during the present observation. We investigated this further and find that there is a minor flare, as seen by the excess H$\alpha$ emission near the middle of the transit (reaching a maximum around the 550th integration). We illustrate this in Figure~\ref{fig:flare}, by showing the average of $F(t) - T(t)F_{\mathrm{out}}$, between integration numbers 520 and 580, where $F(t)$ and $F_{\mathrm{out}}$ correspond to the spectrum time series and the average out-of-transit spectrum, respectively. We find the H$\alpha$ emission to be within one pixel of the expected wavelength of 0.65646 $\mu$m, highlighting the accuracy of the PASTASOSS wavelength calibration. However, the small flare ultimately does not make a significant difference to the final transmission spectrum, since removing the affected integrations does not notably alter the spectrum.

\begin{figure*}
\noindent
\includegraphics[angle=0,width=\textwidth]{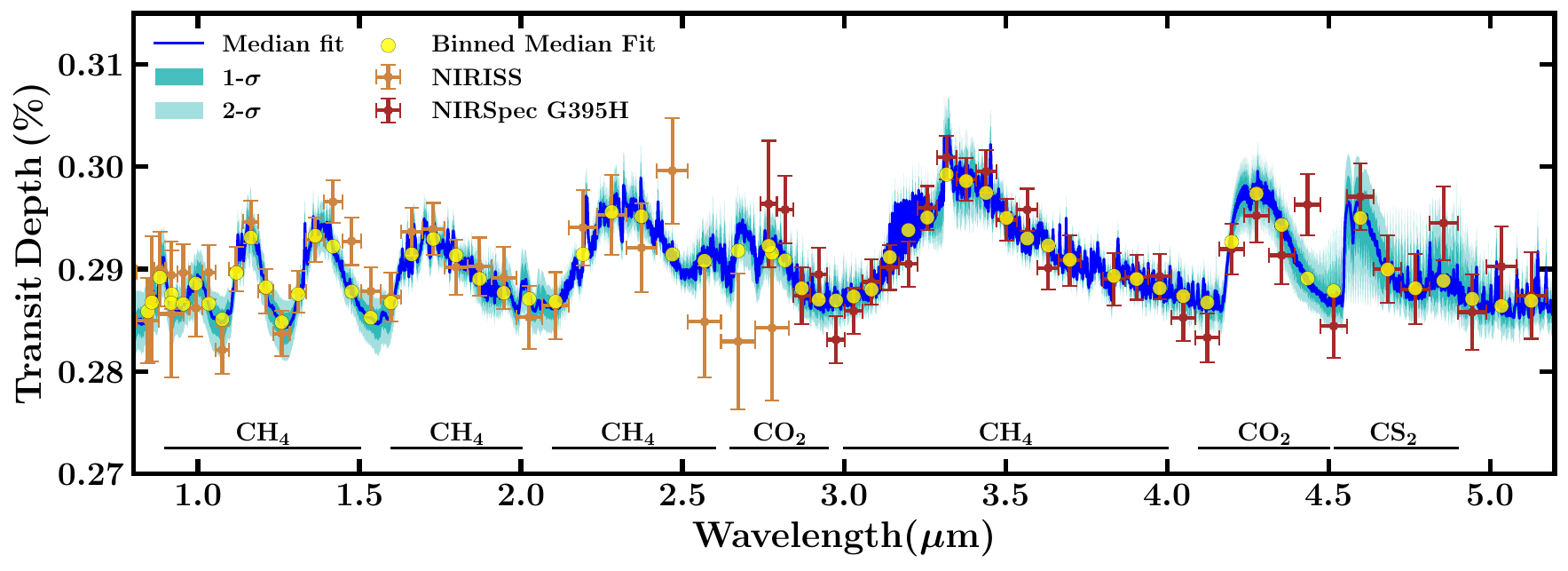}
\caption{JWST transmission spectrum of TOI-270~d. Orange errorbars correspond to NIRISS observations, presented in this work, while dark red errorbars correspond to NIRSpec G395H observations presented by \citetalias{Holmberg2024}. The NIRISS observations were binned down to R=25 for order 1 and R=12 for Order 2 for visual clarity. The blue curve denotes the median retrieved spectral fit obtained with our canonical retrieval, described in Section \ref{sec:methods}. Darker and lighter turquoise shaded regions denote the spectral 1- and 2-$\sigma$ contours corresponding to the spectral fit. Yellow dots denote the median spectrum binned to the resolution of the datapoints. The NIRISS data are vertically offset by +28~ppm, which is the median offset retrieved for our canonical model.}
\label{fig:spectrum}
\end{figure*}

The resulting transmission spectrum is shown in Figures \ref{fig:spectrum} and \ref{fig:spectra_comparison}. It can be seen from Figure \ref{fig:spectra_comparison} that the present reduction of the NIRISS observations is in good agreement with that presented by \citetalias{Benneke2024}. We note that this agreement is especially notable as the two reductions were obtained following different approaches to correcting the 1/f noise.

\begin{table}[h]
\centering
\def\arraystretch{1.2}
\caption{Parameter estimates from the white light curve analysis of the NIRISS SOSS observation of TOI-270~d.}
\vspace{-1mm}
\begin{tabular}{lcccc}
\hline \hline
Parameter & Value\\ \hline
$T_0$ (BJD - 2400000.5 days) & $60346.477984_{-0.000033}^{+0.000031}$ \\ 
$i$ ($^{\circ}$) & $89.756_{-0.037}^{+0.038}$ \\ 
$a / R_*$ & $42.35_{-0.22}^{+0.19}$ \\ 
$u_1$ & $0.117_{-0.036}^{+0.035}$  \\ 
$u_2$ & $0.187_{-0.059}^{+0.062}$ \\ 
$R_\mathrm{p} / R_*$ & $0.05352_{-0.00010}^{+0.00010}$ \\
$R_\mathrm{p}$ ($R_\oplus$) & $2.221 \pm 0.047$ \\ \hline
\end{tabular}
\vspace{-1mm}
\newline
\begin{flushleft}
\footnotesize{\textbf{Note.} The orbital period is held fixed at $11.38014$ days \citep{Evans2023}. We used a stellar radius of $0.380 \pm 0.008$~$R_\odot$ \citep{Kaye2022} to estimate the planet radius.}
\end{flushleft}
\label{tab:wlc_params}
\vspace{-6mm}
\end{table}

\subsection{NIRSpec G395H}
As mentioned above, we used the NIRSpec G395H spectrum of TOI-270~d presented by \citetalias{Holmberg2024} for our canonical retrievals. The observation took place on October 4, 2023, capturing two primary transits of planets~b and~d. The observation utilised the bright-object time series (BOTS) mode along with the F290LP filter, the SUB2048 subarray, and the NRSRAPID readout pattern. In this mode, the spectrum is dispersed onto two different detectors, NRS1 and NRS2, covering around 2.7-5.2~$\mu$m with a small gap between 3.7-3.8~$\mu$m. The observation comprised 1763 integrations, with 11 groups per integration, resulting in a total exposure time of 5.3 hours. From the data reduction cases outlined in \citetalias{Holmberg2024}, we used the dual transit case, where the transits of both planets are fitted simultaneously, providing the most accurate representation of the data. This spectrum covers 2.74-5.17~$\mu$m at a pixel-level resolution. By contrast, \citetalias{Benneke2024} used a lower resolution spectrum which is truncated below 2.86~$\mu$m. Additionally, they discarded three spectral channels due to transit depth outliers or large uncertainties. In \citetalias{Holmberg2024}, they do not report finding or removing any spectral channels with anomalous transit depths.

\section{Atmospheric retrieval methodology}\label{sec:methods}

We used the VIRA retrieval framework \citep{Constantinou2024} to obtain atmospheric constraints from the present observations. VIRA's forward model generator computes transmission spectra via a radiative transfer calculation at the desired spectral resolution. The atmosphere is assumed to be 1-dimensional and in hydrostatic equilibrium. The atmospheric temperature structure is non-uniform and described using the 6-parameter profile of \citet{Madhusudhan2009}. For all models considered, the atmosphere contains H$_2$ and He at solar elemental ratios, which give rise to H$_2$-H$_2$ and H$_2$-He collision-induced absorption \citep{Borysow1988,orton2007,abel2011,richard2012}, setting the spectral baseline. 

We considered atmospheric extinction arising from a number of molecular species. For the common molecules expected in H$_2$-rich atmospheres, namely H$_2$O, CH$_4$, NH$_3$, CO, CO$_2$, H$_2$S and SO$_2$, we used the latest sources of opacity from the POSEIDON database \citep{Macdonald2024}, while the remaining species are adapted from \citet{Madhusudhan2023b}. The opacity sources considered are as follows: H$_2$O \citep{Polyansky2018}, CH$_4$ \citep{Yurchenko2024}, NH$_3$ \citep{Coles2019}, CO \citep{Li2015}, CO$_2$ \citep{Yurckenko2020}, CS$_2$ \citep{Gordon2022} and SO$_2$ \citep{Underwood2016}, C$_2$H$_2$ \citep{Chubb2020}, HCN \citep{Barber2014}, H$_2$S \citep{Azzam2016}, CH$_3$OH \citep{Gordon2022}, C$_2$H$_6$ \citep{Gordon2022}, (CH$_3$)$_2$S (dimethyl sulfide; DMS) \citep{dms_cs2_1,dms_cs2_2, HITRAN2016}, CH$_3$Cl \citep{ch3cl_1, ch3cl_2}, OCS \citep{ocs_1, ocs_2, ocs_3, ocs_4, ocs_5, ocs_6, ocs_7} and N$_2$O \citep{n2o_2}. We note however that some of the cross-sections, for example, DMS, are N$_2$-broadened, which may be somewhat inaccurate for the H$_2$-dominated atmosphere of TOI-270~d.

We begin by reproducing and exploring the findings of \citetalias{Benneke2024} in Section \ref{sec:benneke_reproduction}, who also used the 1-5$\mu$m observations considered in this work. For this, we used an equivalent atmospheric model as the free retrievals of \citetalias{Benneke2024}, which in addition to H$_2$ and He contains H$_2$O, CH$_4$, NH$_3$, CO, CO$_2$, CS$_2$, and SO$_2$. The mixing ratio of each molecule is a free parameter and is uniform across the observable atmosphere. The model also includes opacity contributions from grey clouds and Rayleigh-like hazes, described by a cloud deck pressure $P_\mathrm{c}$ and Rayleigh enhancement factor $a$. We note this model assumes full coverage of the terminator by clouds and hazes and does not fit for the scattering slope, which is set to -4 for Rayleigh scattering. The atmosphere is also assumed to be isothermal, with temperature $T_\mathrm{iso}$. We do not retrieve for an offset between the NIRISS and NIRSpec dataset, as with the canonical model presented by \citetalias{Benneke2024}. Lastly, we highlight that \citetalias{Benneke2024} do not include the reference pressure (or conversely the planetary radius) as a free parameter. We explore how this omission affects the retrieved constraints in Section \ref{sec:benneke_reproduction}, carrying out retrievals with both fixed and free reference pressures. For our own independent retrieval analysis, as presented in Section \ref{sec:canonical_retrievals}, we used the more general VIRA retrieval framework. For this purpose, we considered a broader range of molecules in the model motivated by recent works of temperate sub-Neptunes\citep{Madhusudhan2021, Madhusudhan2023b, Holmberg2024}.

Our canonical model atmosphere includes spectral contributions from clouds and hazes following the parametric approach of \citet{Pinhas2018}. This approach comprises of a grey opacity cloud deck at a parametric cloud top pressure $P_\mathrm{c}$, and Rayleigh-like haze scattering described by an enhancement factor $a$ and slope parameter $\gamma$. The clouds and hazes can have a patchy coverage of the observed terminator atmosphere, described by the coverage fraction $f_\mathrm{c}$. The reference pressure is a free parameter for all retrievals.

We set the planetary radius to coincide with the planet-star radius ratio obtained from the NIRSpec white light curve. Consequently, the retrieved reference pressure effectively constitutes an estimate of the effective photospheric pressure across the NIRSpec range \citep{Constantinou2024}.

For all our retrievals we also retrieved an offset between the NIRISS and NIRSpec datasets as pursued in previous work \citep{Madhusudhan2023b}. Moreover, all retrievals compute the goodness of fit of atmospheric models by considering the empirical covariance matrix of the NIRISS observation, following \citet{Constantinou2024}. We explore the available parameter space using the MultiNest Bayesian nested sampling implementation \citep{Buchner2014}. 

For our canonical retrievals, the Bayesian priors used for the molecular mixing ratios are log-uniform between 10$^{-12}$-10$^{-0.3}$. The prior for $T_0$, the temperature at the top of the atmosphere, is uniform between 50-600~K. The priors and descriptions of all parameters in the canonical model are detailed in Table \ref{tab:retrieval_priors}.

\begin{figure*}
\noindent
\includegraphics[angle=0,width=\textwidth]{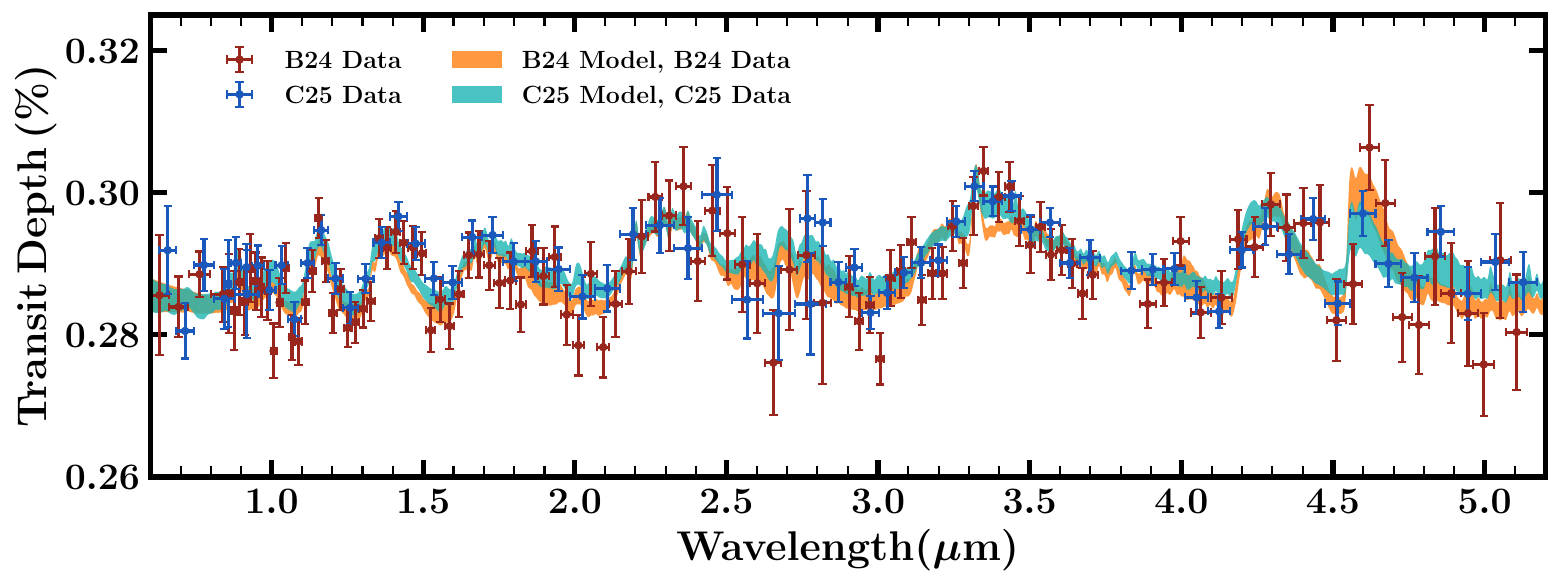}
\caption{JWST transmission spectra of TOI-270~d presented by \citetalias{Benneke2024} (red errorbars) and those used for our canonical retrievals (blue errorbars), combining NIRSpec data by \citetalias{Holmberg2024} with the NIRISS observation reduced in this work. Also shown are the 1-$\sigma$ contours of the retrieved spectral fits to either dataset. The orange contours correspond to the retrieved spectral fit to the red errorbars using an equivalent atmospheric model to the free chemistry retrievals of \citetalias{Benneke2024}, as discussed in Section \ref{sec:benneke_reproduction}. The turquoise contours correspond to the spectral fit to the blue errorbars using our canonical model, as discussed in Section \ref{sec:canonical_retrievals}. The present (blue) NIRISS data has been vertically offset by +28~ppm, the median offset retrieved for our canonical model, as with Figure \ref{fig:spectrum}. C25 in the figure caption refers to the data and canonical model presented in this work.}
\label{fig:spectra_comparison}
\end{figure*}

We computed the detection significance for a number of molecular species using Bayesian model comparison. Specifically, for each molecular species, we carry out an additional retrieval with that molecule excluded from the model. We then compare the model evidence computed for this restricted model with that obtained for our canonical model, obtaining the molecule's detection significance expressed as a frequentist sigma-significance \citep{Sellke2001, trotta2008, benneke2013}. We note that with this approach, the sigma-significance values quoted represent maximal values. We quote the sigma-significance of each model comparison with an error, which arise from the uncertainty estimation provided by MultiNest for the evidence computed for each of the two models under consideration. As a robustness check, we carry out an additional retrieval with UltraNest \citep{Buchner2021}, which trades computational efficiency for more rigorous sampling of the prior space, to assess the accuracy of the computed Bayesian evidence values, which we detail in Appendix \ref{sec:robustness_checks}. For all constraints quoted in this work, the central value corresponds to the median of the retrieved posterior distribution which corresponds to the 50$^\mathrm{th}$ percentile, while the upper and lower 1-$\sigma$ values correspond to the 84$^\mathrm{th}$ and 16$^\mathrm{th}$ percentile, respectively. Quoted 2-$\sigma$ upper limits meanwhile correspond to the 95$^\mathrm{th}$ percentile of the given distribution. We note that in cases where the posterior distribution is not well constrained or asymmetric this approach may not necessarily include the mode in the above intervals. We also present equivalent constraints computed through the highest posterior density interval approach in Table \ref{tab:retrieval_priors} in the Appendix.

Beyond the above canonical model, we also considered Mie scattering aerosols, which replace the above parametric cloud and haze model. Following \citet{Constantinou2023} and \citet{Pinhas2017}, we carry out Mie scattering calculations for aerosols consisting of H$_2$O in ice \citep{Warren2008} and liquid \citep{Segelstein1981} states, which are the result of condensation, as well as Titan-like tholins\footnote{We specifically use the refractive indices for 300~K, which corresponds to the approximate temperature retrieved at the photosphere with our canonical retrievals.} \citep{He2024} which can arise from photochemical processes. The free parameters of this model consist of the mixing ratio of each aerosol species, the global modal particle size $r_\mathrm{c}$, the aerosol scale height, $H_\mathrm{c}$ and the terminator coverage fraction $f_\mathrm{c}$.

Lastly, we also conducted retrievals which consider unocculted stellar heterogeneities in addition to the canonical atmospheric model. Stellar heterogeneities beyond the transit chord result in the spectrum of light passing through the atmosphere not being the same as the unocculted disk-averaged spectrum, which contains contributions from the pristine photosphere and heterogeneities. As a result, unocculted starspots and faculae can effectively distort the observed atmospheric transmission spectrum. For this work, we included stellar heterogeneities following the prescription of \citet{Rackham2017} and \citet{Pinhas2018}. It comprises of three free parameters: the effective temperature of the pristine photosphere, the effective temperature of the averaged heterogeneities and the fraction covered by stellar heterogeneities. We used PHOENIX spectra \citep{Husser2013} for the photosphere and heterogeneities, fixing the stellar properties other than temperature to literature values. We used uninformative, uniform priors ranging between 0-0.5 for the coverage fraction and between 1800-4900~K for the temperature of the heterogeneities. For the photospheric temperature, we used a Gaussian prior informed by literature measurements of the stellar effective temperature \citep{van2021masses}.

\begin{figure*}
\noindent
\includegraphics[angle=0,width=\textwidth]{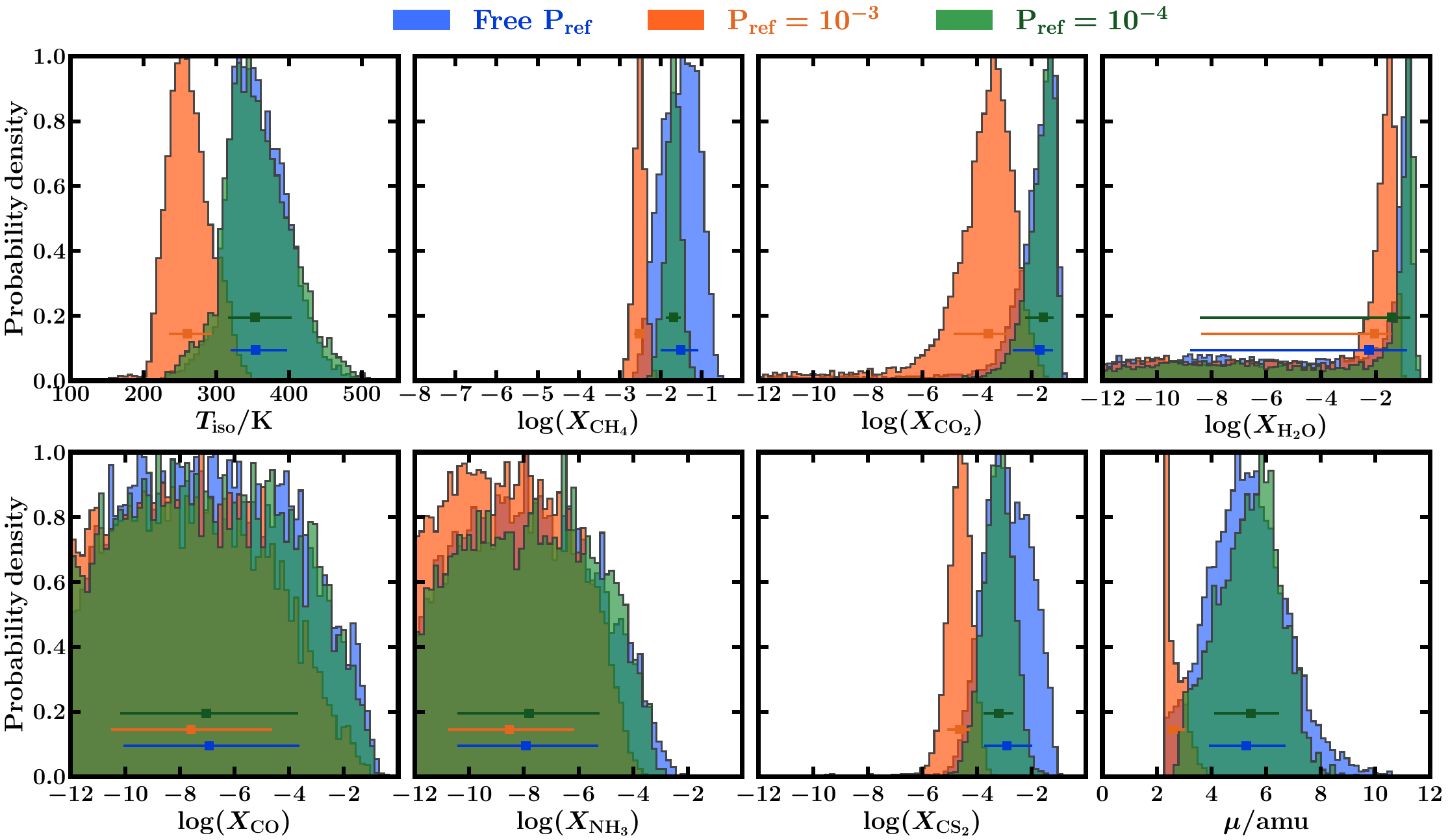}
\caption{Atmospheric parameter constraints obtained with retrievals on JWST NIRISS and NIRSpec observations presented by \citetalias{Benneke2024}, using an equivalent free chemistry atmospheric model to \citetalias{Benneke2024}. The blue posterior distributions correspond to the retrieval where the reference pressure, $P_\mathrm{ref}$ is a free parameter, while the orange posterior distributions correspond to a retrieval with $P_\mathrm{ref}$ fixed to $10^{-3}$~bar, and the green to a retrieval with $P_\mathrm{ref} = 10^{-4}$~bar. $\mu$ denotes the MMW of the atmosphere, which is inferred from the mixing ratio posterior distributions of all molecules in the model. Errorbars denote the median and 1-$\sigma$ interval, while arrows denote 2-$\sigma$ upper estimates.}
\label{fig:benneke_reproduction}
\end{figure*}

\section{Results: Assessment of mixed envelope scenario}
\label{sec:benneke_reproduction}

As noted above, \citetalias{Benneke2024} carried out a retrieval analysis of the same NIRISS and NIRSpec G395H observations with an independent reduction. \citetalias{Benneke2024} find strong evidence for the presence of CH$_4$ and CO$_2$ in the atmosphere of TOI-270~d, and less confident indications of CS$_2$ and H$_2$O. CH$_4$, CO$_2$ and H$_2$O are constrained to log-mixing ratios of $-1.64^{+0.38}_{-0.36}$,  $-1.67^{+0.40}_{-0.60}$ and $-1.10^{+0.31}_{-0.92}$, which are high enough to impact the atmospheric MMW. CS$_2$ meanwhile is constrained to a log-mixing ratio of $-3.44^{+0.66}_{-0.67}$. Based on these retrieved mixing ratios, \citetalias{Benneke2024} find an inferred MMW constraint of $5.47^{+1.25}_{-1.14}$~amu. Their retrievals also find a relatively high isothermal temperature constraint of $385^{+44}_{-41}$~K with free chemistry retrievals. Together, these constraints led to the conclusion that TOI-270~d is a miscible-envelope sub-Neptune,possessing a deep, well-mixed atmosphere whose high mean molecular weight is offset by a relatively high temperature, which can give rise to prominent atmospheric absorption features.

In a separate analysis, \citetalias{Holmberg2024} also found the same molecular detections of CH$_4$, CO$_2$, and CS$_2$ solely considering the NIRSpec data, combined with prior observations with HST WFC3. However, \citetalias{Holmberg2024} constrain the mixing ratios of CH$_4$, CO$_2$, and H$_2$O to values that are $\sim$0.5-1~dex lower than the equivalent constraints of \citetalias{Benneke2024}. Specifically, in the dual-transit retrieval case, \citetalias{Holmberg2024} constrain CH$_4$ to a log-mixing ratio of $-2.72^{+0.41}_{-0.50}$, that of CO$_2$ to $-2.46^{+0.71}_{-0.92}$ and H$_2$O to $-1.91^{+0.57}_{-0.94}$. These lower atmospheric abundances correspond to lower MMW values than those of \citetalias{Benneke2024}. At the same time, \citetalias{Holmberg2024} constrain CS$_2$ to a mixing ratio of $-3.07^{+0.74}_{-0.91}$ which is $\sim$0.4 dex higher than that obtained by \citetalias{Benneke2024} but consistent to within 1-$\sigma$. Moreover, \citetalias{Holmberg2024} also constrain the terminator temperature at 10~mbar, corresponding to the observable atmosphere, to $289^{+80}_{-75}$~K, which is less precise but overall cooler than that of \citetalias{Benneke2024}.

In this section we reproduce the findings of \citetalias{Benneke2024} using the VIRA retrieval framework, determining the modelling choices that can lead to the above constraints. We note that in addition to the above constraints obtained with free chemistry retrievals \citetalias{Benneke2024} obtained equivalent and consistent constraints using a quenched-chemistry retrieval. We focus on the findings of the free chemistry retrieval to enable a direct comparison. We carry out this analysis on both the data presented by \citetalias{Benneke2024}, as well as the combination of NIRSpec data presented by \citetalias{Holmberg2024} and NIRISS presented in Section \ref{subsec:niriss_reduction}, to assess whether the retrieved constraints vary depending on the reduction pipelines used. For both sets of reductions, we consider retrievals with a fixed reference pressure (and planetary radius), as done by \citetalias{Benneke2024} as well as retrievals where it is a free parameter. Beyond the reference pressure, we used similar model assumptions as \citetalias{Benneke2024}, as described in Section \ref{sec:methods}. We note that the present modelling may have minor differences from that employed by \citetalias{Benneke2024}, arising from, for example, the molecular opacities used.

\subsection{Atmospheric model of \citetalias{Benneke2024}}

As discussed in Section \ref{sec:methods}, the free chemistry atmospheric model used by \citetalias{Benneke2024} consists of an H$_2$ and He background atmosphere which also includes H$_2$O, CH$_4$, CO, CO$_2$, NH$_3$, CS$_2$, and SO$_2$, with the mixing ratios of the seven molecular species being free parameters. The terminator atmosphere is assumed to be isothermal, parametrised by one temperature free parameter. Atmospheric clouds and hazes are described by another two free parameters, comprising of the cloud top pressure and the Rayleigh enhancement factor. In total the model consists of 10 free parameters as noted by \citetalias{Benneke2024}. We highlight that this model does not treat the reference pressure or, equivalently, the planetary radius, as a free parameter. Additionally, the clouds and hazes model only considers Rayleigh-like scattering slopes with $\lambda^{-4}$, and assumes that the terminator is fully covered by clouds and hazes, if present.

\subsection{Using \citetalias{Benneke2024} data}
\label{subsec:bennekedata_bennekemodel}

 Figure \ref{fig:spectra_comparison} shows the retrieved spectral fit we obtain applying the atmospheric model of \citetalias{Benneke2024} to the NIRISS and NIRSpec data presented by \citetalias{Benneke2024}, with reference pressure left as a free parameter. Figure \ref{fig:benneke_reproduction} shows the corresponding retrieved posterior distributions for the atmospheric temperature and molecular mixing ratios. Fixing the reference pressure to $10^{-4}$~bar leads to atmospheric constraints consistent with those obtained by \citetalias{Benneke2024}. The log-mixing ratio of CH$_4$ is constrained to $-1.68^{+0.17}_{-0.19}$, while that of CO$_2$ is constrained to $-1.58^{+0.37}_{-0.69}$. The posterior distribution of H$_2$O is peaked at high abundances pushing against the upper end of the mixing ratio prior. As a result, the overall inferred MMW is $5.43^{+1.04}_{-1.33}$~amu, which is consistent with that obtained by \citetalias{Benneke2024}. The retrieval also obtains constraints for CS$_2$ at $-3.21^{+0.54}_{-0.56}$, and also obtained a peaked but largely unconstrained posterior for SO$_2$, with a mode at $\sim$-4 dex log-mixing ratio. Lastly, the isothermal temperature is constrained to $353^{+51}_{-37}$ K which is somewhat lower than that obtained by \citetalias{Benneke2024} but consistent within the 1-$\sigma$ uncertainties.

 Fixing the reference pressure to $10^{-3}$~bar leads to lower overall atmospheric abundances. CH$_4$ and CO$_2$ are constrained to log-mixing ratios of $-2.51^{+0.13}_{-0.14}$ and $-3.55^{+0.80}_{-1.26}$, respectively, while CS$_2$ is constrained to $-4.64^{+0.41}_{-0.46}$. The posterior distribution for H$_2$O displays a more prominent tail towards lower abundances, which leads to a lower overall inferred MMW of $2.58^{+0.21}_{-0.47}$~amu, consistent with an H$_2$ and He dominated atmosphere. The isotherm temperature is also lower, at 260$^{+34}_{-25}$~K, which serves to compensate for the lower MMW of the atmosphere.

 We note that these constraints are sensitive to the specific choice of $P_\mathrm{ref}$. For instance, fixing $P_\mathrm{ref}$ to $10^{-4.5}$~bar leads to a log-mixing ratio constraint of $-1.14^{+0.13}_{-0.13}$ for CH$_4$, which is $\sim$0.5 higher than that retrieved with $P_\mathrm{ref} = 10^{-4}$~bar above. Similarly, the log-mixing ratio of CO$_2$ is constrained to $-1.57^{+0.40}_{-1.98}$ and that of CS$_2$ to $-2.20^{+0.50}_{-0.53}$. The H$_2$O posterior meanwhile is primarily unconstrained with only a slight peak towards the upper end of the prior. The temperature meanwhile is constrained to $351^{+43}_{-31}$~K.

 Leaving the reference pressure as a free parameter, CH$_4$ is constrained to a log-mixing ratio of $-1.50^{+0.54}_{-0.52}$ and CO$_2$ to $-1.71^{+0.49}_{-1.00}$. CS$_2$ meanwhile is constrained to $-2.90^{+0.93}_{-0.85}$, while SO$_2$ displays the same peaked but unconstrained posterior as in the above cases, with a mode at $\sim$-4 log-mixing ratio. The isothermal temperature is constrained to $354^{+43}_{-34}$K. Lastly, H$_2$O again has a posterior with a peak at high abundances, but this time has a much more pronounced tail towards low abundances. As a result, this retrieval leads to an overall inferred MMW of $5.27^{+1.37}_{-1.44}$, which is similar to that obtained from $P_\mathrm{ref}=10^{-4}$~bar case. As such these constraints are also consistent with those of \citetalias{Benneke2024} within the 1-$\sigma$ intervals.

We therefore find that we can reproduce the constraints obtained by \citetalias{Benneke2024} both with and without $P_\mathrm{ref}$ as a free parameter. In the following sections we investigate whether these constraints are dependent on which data reduction considered and whether they are affected by changes to the atmospheric model.

\subsection{Using present data}
\label{subsec:presentdata_bennekemodel}

We now consider retrievals using the above atmospheric model of \citetalias{Benneke2024} with the NIRSpec reduction of \citetalias{Holmberg2024} and NIRISS reduction presented in Section \ref{subsec:niriss_reduction}. In doing so, we assess whether the constraints obtained by \citetalias{Benneke2024} and in the above section are affected by the choice of specific reduction of the same observations. We note that for this reduction of the data we consider linear offsets between the NIRISS and NIRSpec observations as there is no prior indication that none are needed.

Maintaining $P_\mathrm{ref}$ as a free parameter, we find constraints that are consistent with those obtained above. In particular, we retrieve a CH$_4$ log-mixing ratio constraint of $-1.62^{+0.30}_{-0.27}$, and a CO$_2$ constraint of $-1.47^{+0.29}_{-0.66}$. We once again obtain an H$_2$O posterior distribution with a peak towards the upper end of the prior but with a substantial tail towards lower abundances. Similarly SO$_2$ displays a peaked posterior at $\sim$-4 log-mixing ratio but also with a substantial low abundance tail. The temperature meanwhile is constrained to $304^{+43}_{-53}$~K which is lower but still consistent with that obtained in the prior section. The abundance constraint for CS$_2$ is the only notable difference, with this retrieval obtaining a constraint of $-4.99^{+0.98}_{-2.55}$.

We note that studies have shown that atmospheric retrievals can be highly sensitive to minor differences in the data arising from different reduction pipelines both for JWST \citep{Constantinou2023} and prior facilities \citep[e.g.][]{Himes2022}. As such, the agreement between the present results with two different datasets is encouraging, and suggest that pipeline differences are not the primary contributor to the different abundance constraints obtained by prior studies. In the following section, we consider how changes to the atmospheric model can impact the retrieved constraints.

\subsection{Dependence on atmospheric model}
\label{subsec:bennekedata_modelcomparison}

\begin{figure}
\noindent
\includegraphics[angle=0,width=\columnwidth]{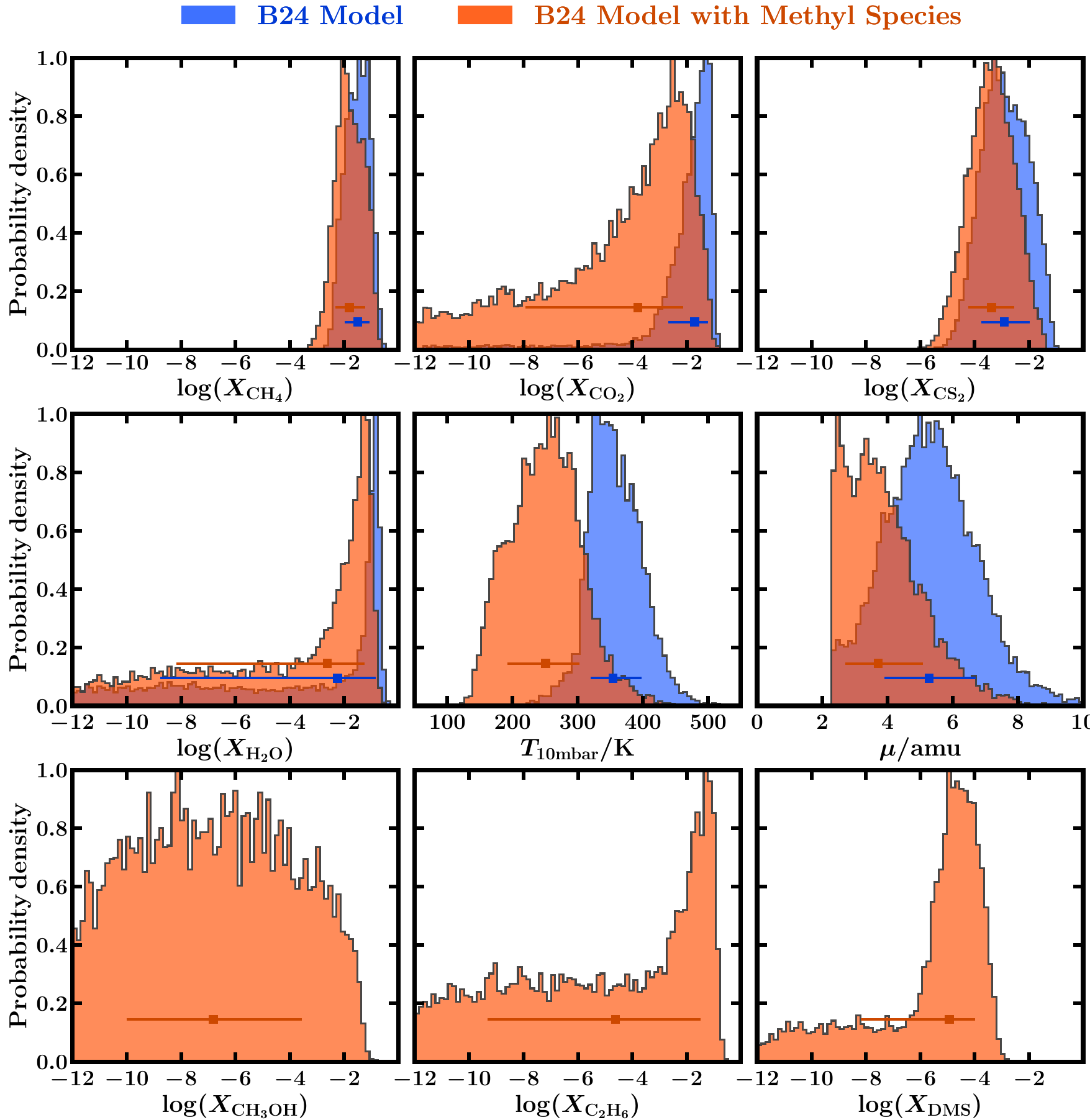}
\caption{Posterior distributions obtained by atmospheric retrievals on the JWST spectrum from \citetalias{Benneke2024}. The blue distributions correspond to a retrieval using the an equivalent free chemistry atmospheric model to \citetalias{Benneke2024}. The orange distributions were obtained with the atmospheric model of \citetalias{Benneke2024} but additionally including spectral contributions from CH$_3$OH, C$_2$H$_6$, and DMS. }
\label{fig:benneke_methylspecies}
\end{figure}

Having established that the retrieved atmospheric constraints persist across different reductions of the observations, we now consider how they are impacted by the specific choice of atmospheric model. Using the data reduction presented by \citetalias{Benneke2024}, we compare the constraints obtained with the model of \citetalias{Benneke2024}, with those obtained when we additionally include the methyl-bearing species C$_2$H$_6$, CH$_3$OH, and DMS in the model. C$_2$H$_6$ and DMS were also considered by \citetalias{Holmberg2024} and we additionally include CH$_3$OH.

As shown in Figure \ref{fig:benneke_methylspecies}, the inclusion of methyl-bearing species affects the retrieved atmospheric properties. Specifically,  CH$_4$ is constrained to a log-mixing ratio of $-1.81^{+0.59}_{-0.51}$, that of CO$_2$ to $-3.81^{+1.69}_{-4.17}$ and CS$_2$ to $-3.37^{+0.85}_{-0.87}$. The isothermal terminator temperature meanwhile is constrained to $251^{+52}_{-59}$~K. Lastly, the inferred MMW is $3.79^{+1.37 }_{-1.10}$ amu.

We also obtain peaked posterior distributions for the mixing ratios of the newly added DMS and C$_2$H$_6$. In particular, the DMS posterior shows a prominent peak with a tail towards lower abundances, while that of C$_2$H$_6$ is somewhat peaked. CH$_3$OH remains largely unconstrained.  All three species give rise to similar absorption near 3.5$\mu$m, which is needed to explain the observations. Beyond impacting the retrieved atmospheric properties, the addition of the methyl-bearing species also provide a better explanation for the data. Comparing the model evidence values computed for the basic \citetalias{Benneke2024} model and for the model including the methyl species, we find that there is a preference for the latter at a 2.84 $\sigma$ level. This indicates that the present observations show signs excess of atmospheric absorption by methyl-bearing species beyond CH$_4$, but are insufficient to determine which species in particular is responsible.

The retrieved CH$_4$ abundance and temperature constraints are consistent with those obtained with our canonical retrieval model in Section \ref{sec:canonical_retrievals}. Additionally, the retrieved CH$_4$ log-mixing ratio is between that retrieved by \citetalias{Holmberg2024} ($-2.72^{+0.41}_{-0.50}$) and that retrieved by \citetalias{Benneke2024} ($-1.64^{+0.38}_{-0.36}$) and consistent with both to within 1-$\sigma$. The CS$_2$ constraint is lower than that retrieved by \citetalias{Holmberg2024} ($-3.07^{+0.74}_{-0.91}$) and \citetalias{Benneke2024} ($-3.44^{+0.66}_{-0.67}$) but again within 1-$\sigma$ of both. Finally, the retrieved isothermal temperature is also lower but still within 1-$\sigma$ the value of $289^{+80}_{-75}$~K obtained by \citetalias{Holmberg2024} (at 10~mbar) and more than 1-$\sigma$ lower than that retrieved by \citetalias{Benneke2024} at $385^{+44}_{-42}$~K.

We conclude that the atmospheric properties retrieved by \citetalias{Benneke2024} were influenced by the specific choice of atmospheric model with a limited set of atmospheric species. The present findings highlight the importance of considering a broad set of physically plausible chemical species in atmospheric retrievals, as the omission of even trace species that can contribute minor but observable spectral features can potentially bias the retrieved abundances.

\section{Results: Atmospheric retrieval}\label{sec:canonical_retrievals}

\begin{table*}[]
    \centering
    \addtolength{\tabcolsep}{-0.5em}
    \def\arraystretch{1.5}
    \caption{Atmospheric composition and temperature constraints obtained for different models used for the retrievals.}
    \vspace{-1mm}
    \begin{tabular}{l||ccccccc|c|c}
        Model & CH$_4$ & CO$_2$ & H$_2$O & CS$_2$ & C$_2$H$_6$ & DMS  & NH$_3$ & $T_\mathrm{10mbar}$/K & log($\mathcal{Z}$) \\
        \hline
        \hline
        Canonical & $-1.86^{+0.30}_{-0.29}$ & $-1.71^{+0.38}_{-0.66}$ & $-1.88^{+0.78}_{-4.13}$ & $-4.74^{+0.65}_{-1.10}$ & $-2.20^{+0.87}_{-2.01}$ & ($-4.58$) & ($-4.24$) & $323^{+58}_{-52}$ & 23698.54\\
         & $6.36\sigma$ & $3.90\sigma$ & $2.11\sigma$  & $2.04\sigma$ & $2.14\sigma$ & - & - & -\\
        \hline
        No C$_2$H$_6$ & $-1.77^{+0.32}_{-0.30}$ & $-1.89^{+0.50}_{-0.77}$  & $-1.52^{+0.53}_{-4.07}$ & $-4.81^{+0.65}_{-1.14}$ & -  & $-5.80^{+1.00}_{-3.71}$ & ($-4.08$) & $335^{+58}_{-49}$ & 23697.35\\
        No C$_2$H$_6$, DMS & $-1.70^{+0.32}_{-0.31}$ & $-1.82^{+0.47}_{-0.76}$ & $-1.64^{+0.64}_{-5.25}$ & $-4.78^{+0.65}_{-1.42}$ & - & - & ($-1.51$) & $256^{+62}_{-72}$ & 23696.67\\

        \hline 
        Liquid H$_2$O Aerosols & $-2.17^{+0.37}_{-0.50}$ & $-1.73^{+0.41}_{-0.83}$ & $-2.15^{+0.90}_{-1.25}$  & $-4.87^{+0.68}_{-0.92}$ & $-1.94^{+0.86}_{-2.14}$ & ($-4.99$)  & ($-4.58$) & $302^{+56}_{-59}$ & 23697.38\\
        Ice H$_2$O Aerosols & $-2.21^{+0.34}_{-0.46}$ & $-1.74^{+0.41}_{-0.79}$ & $-2.22^{+0.92}_{-1.26}$  & $-4.88^{+0.71}_{-0.94}$ & $-1.79^{+0.72}_{-1.46}$ & ($-5.15$) & ($-4.61$)& $295^{+60}_{-58}$ & 23697.32\\
        Tholin Aerosols & $-2.05^{+0.29}_{-0.34}$ & $-1.62^{+0.33}_{-0.57}$ & $-1.77^{+1.67}_{-1.20}$  & $-4.75^{+0.65}_{-0.91}$ & $-2.28^{+1.07}_{-3.73}$ & ($-4.93$) & ($-4.34$) & $313^{+59}_{-59}$  &  23698.22\\
        \hline
    \end{tabular}
    \vspace{1mm}
    \newline
    \begin{flushleft}
    \footnotesize{ \textbf{Note.} The models include our canonical atmospheric model, more restricted models with different combinations of C$_2$H$_6$ and DMS present, and models with Mie-scattering aerosols instead of parametric clouds and hazes. Also shown are the individual detection significances for the constrained species in the canonical model. The molecular abundances are shown as log-mixing ratios. Values in brackets denote the 2-$\sigma$ upper limit, corresponding to the 95$^\mathrm{th}$ percentile. Dashes indicate molecules which were not present in a particular model.}
    \end{flushleft}
    \label{tab:constraints}
\end{table*}

We now present a retrieval analysis of the current data, including the NIRISS data described in Section \ref{sec:observations} combined with the NIRSpec G395H data reported by \citetalias{Holmberg2024}, as shown in Figure \ref{fig:spectrum}. We use a more elaborate atmospheric model than that used in Section \ref{sec:benneke_reproduction}, which includes a non-isothermal temperature profile, variable haze scattering slopes and partial clouds and hazes coverage of the terminator, as detailed in Section \ref{sec:methods}. We establish the detection significance for a molecular species by comparing the Bayesian evidence of our canonical model with one where the molecule (or group of molecules) is absent. The retrieved abundance constraints of key molecular species, as well as the atmospheric temperature at 10~mbar, are summarised in Table \ref{tab:constraints}.

\subsection{Atmospheric constraints}\label{subsec:canonical_constraints}

The retrieved spectral fit is shown in Figure \ref{fig:spectrum}. The spectrum contains several CH$_4$ absorption features that can be visually identified between $\sim$0.6-4$\mu$m, while CO$_2$ and CS$_2$ give rise to the spectral features seen at $\sim$4.3 and 4.7~$\mu$m, respectively. CO$_2$ also contributes a less prominent absorption feature at $\sim$2.7~$\mu$m. As a result of these prominent absorption features, the retrieval constrains the log-mixing ratio of CH$_4$ to $-1.86^{+0.30}_{-0.29}$, that of CO$_2$ to $-1.71^{+0.38}_{-0.66}$ and CS$_2$ to $-4.74^{+0.65}_{-1.10}$, as seen in Figure \ref{fig:canonical_posteriors}. H$_2$O meanwhile is constrained to a log-mixing ratio of $-1.88^{+0.78}_{-4.13}$, with a posterior distribution displaying a tail towards lower abundances. The retrieved posterior distributions for key molecular species and the photospheric temperature are shown in Figure \ref{fig:canonical_posteriors}. The retrieved posteriors and constraints for all parameters in the canonical model are presented in Figure \ref{fig:canonical_posteriors_full} and Table \ref{tab:retrieval_priors}, respectively. Additionally, the canonical retrieval also places a low upper-limit on the abundance of NH$_3$, with a 2-$\sigma$ upper limit of -4.24, suggesting a very low probability to NH$_3$ being present in abundances approaching that of CH$_4$. CO meanwhile is also unconstrained, with a 2-$\sigma$ log-mixing ratio upper limit of $-1.32$. Lastly, the posterior distribution of SO$_2$ is unconstrained, with a 2-$\sigma$ upper limit of $-4.52$.

The retrieval additionally finds indications for the presence of methyl-bearing species, which provide excess opacity near 3.5~$\mu$m, beyond that of CH$_4$ alone. The log-mixing ratio of C$_2$H$_6$ is constrained to $-2.20^{+0.87}_{-2.01}$ with a tail towards low abundances. As noted above C$_2$H$_6$ provides additional absorption at $\sim$3.5~$\mu$m, which is potentially degenerate with other similar methyl-bearing species such as DMS. DMS also shows a peak in its otherwise unconstrained posterior distribution. We explore this excess methyl-group absorption further in Section \ref{subsec:detections}.

\subsubsection{Atmospheric temperature}
\begin{figure*}
\noindent

\includegraphics[angle=0,width=\textwidth]{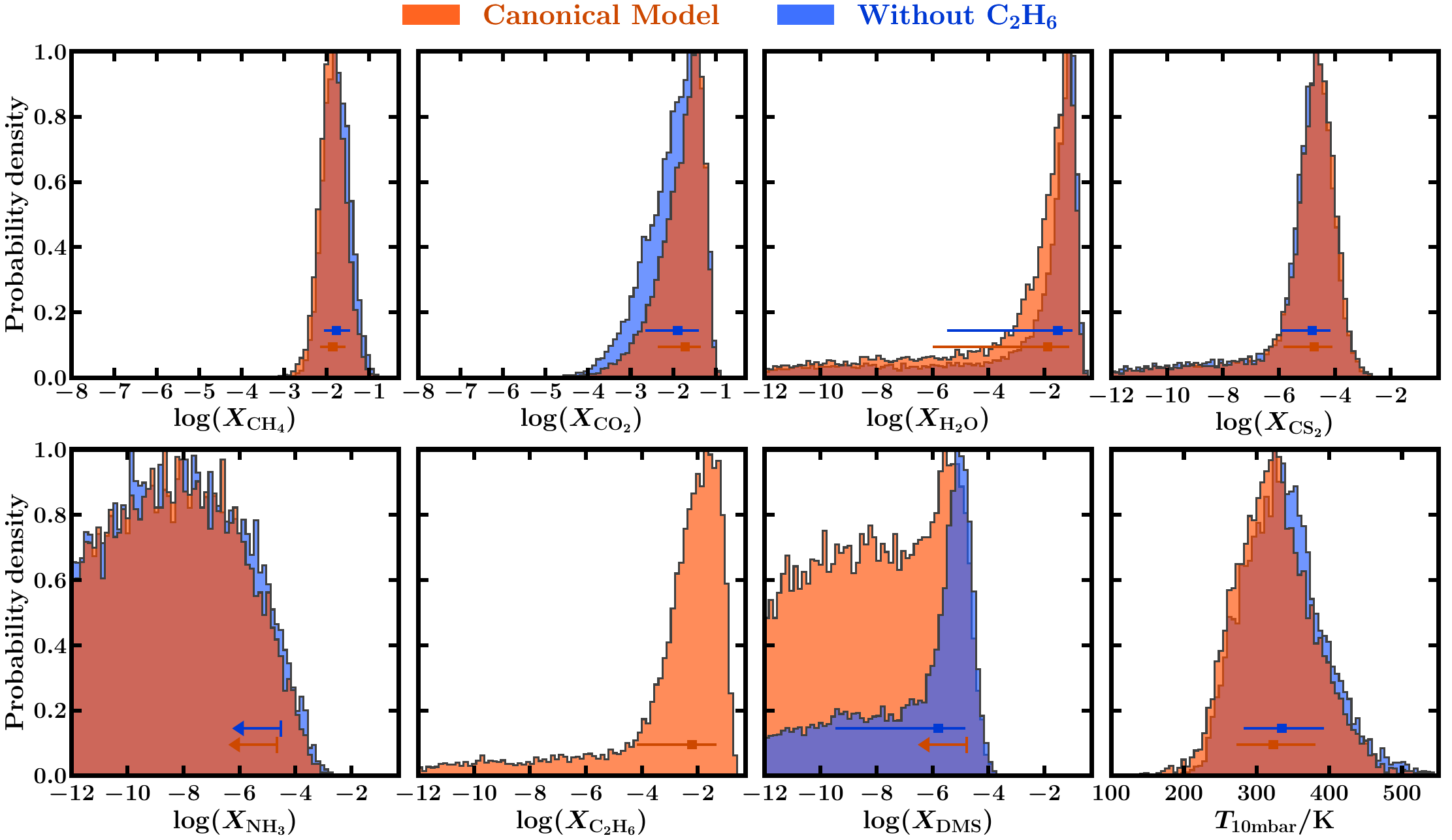}
\caption{Posterior probability distributions of key molecular species and photospheric temperature retrieved with the canonical atmospheric model described in Section \ref{sec:methods}, shown in orange. Also shown in blue are the corresponding results with C$_2$H$_6$ removed from the model, whereby a peak in the DMS posterior is found. Horizontal points with errorbars denote the median and 1-$\sigma$ interval of the distribution of the same colour, while arrows denote 2-$\sigma$ upper limits. }
\label{fig:canonical_posteriors}
\end{figure*}

The atmospheric retrieval constrains the temperature at the top of the atmosphere at 1 $\mu$bar, to $273^{+51}_{-48}$~K. At a pressure of 10~mbar, which is within atmospheric region typically probed by transmission spectroscopy \citep{Constantinou2024}, the retrieved temperature is $323^{+58}_{-52}$~K as shown in Figure \ref{fig:canonical_posteriors}. The retrieved P-T profile shown in Figure \ref{fig:canonical_posteriors_full} gradually increases with temperature towards lower altitudes, but remains consistent with an isotherm to within 1-$\sigma$. 

\subsubsection{Clouds and hazes}
In addition to the above constraints on atmospheric composition and temperature, we also retrieve constraints for the cloud deck pressure $\mathrm{log(}P_\mathrm{c} / \mathrm{bar}) = -2.09^{+0.50}_{-0.48}$. The clouds and hazes coverage fraction is constrained towards full coverage, that is, pushing against the upper end of the prior range. The Rayleigh-like haze scattering parameters meanwhile are largely unconstrained. Together, the above constraints correspond to some truncation of the observed spectral features across the wavelength range, while still permitting the significant absorption peaks that gave rise to the molecular constraints detailed above.

\subsection{Molecular detections}\label{subsec:detections}

As detailed in Section \ref{sec:methods}, we obtain the detection significance for each molecular species by carrying out additional retrievals, each of which exclude one molecule, and comparing their model evidence to that of the canonical model. In this Section, we report on the computed detection significance values for each of the molecules that were constrained in Section \ref{sec:canonical_retrievals}. Additionally, in the case of the methyl-bearing species C$_2$H$_6$ and DMS, we also report on how their inclusion or exclusion from the model affects the abundances of other molecular species and other atmospheric properties.

\subsubsection{Spectrally prominent species}

For the spectrally prominent species CH$_4$ and CO$_2$, we find detection significance values indicating high confidence detections. CH$_4$ is detected with a high significance of  $6.36 \pm 0.03 \sigma$, rendering it the most confidently detected species for the present observations. CO$_2$ is detected at a lower but still confident significance of  $3.90 \pm 0.06 \sigma$, indicating a robust detection. CS$_2$ meanwhile is detected with a significance of $2.04 \pm 0.10~\sigma$, while H$_2$O is detected at a $2.11 \pm 0.08~\sigma$ level. We do not find significant evidence for any other potentially prominent species such as NH$_3$, CO, H$_2$S or SO$_2$.

\subsubsection{Methyl-bearing species}

Besides CH$_4$, we find nominal evidence for other methyl-bearing species. Using our canonical model, we find C$_2$H$_6$ at a low detection significance, at $2.1~\pm0.1 \sigma$. This is because its spectral contributions are readily replaceable by other comparable methyl-bearing species. This is indicated by the slight peak in the DMS posterior. Consequently, on excluding C$_2$H$_6$ the retrieval invokes DMS to explain the same spectral features, resulting in a more prominently peaked DMS posterior. This replacement still results in a good fit to the data, as evidenced by the tentative detection significance of C$_2$H$_6$ alone. Moreover, this result is consistent with our findings detailed in Section \ref{subsec:bennekedata_modelcomparison}, where we obtain a peaked DMS posterior when including methyl-bearing species in the model of \citetalias{Benneke2024} using data presented by \citetalias{Benneke2024}. Comparing the canonical model, which includes both C$_2$H$_6$ and DMS, to one that excludes both species, we find a preference at a $2.5 \pm 0.1 \sigma$ level.

Our findings therefore indicate the presence of excess absorption at $\sim$3.5$\mu$m beyond that of CH$_4$ by other methyl-bearing species. These findings are consistent with the model preference we find in favour of including methyl-bearing species when considering the data and atmospheric model of \citetalias{Benneke2024} presented in Section \ref{subsec:bennekedata_modelcomparison}. We therefore conclude that the present observations indicate that methyl bearing species are present at a confidence equivalent to or greater than that for CS$_2$ and H$_2$O, but are insufficient to determine which species or group of species is specifically present.

\subsection{Mean molecular weight}
The abundance constraints obtained with our canonical retrieval lead to an inferred MMW constraint of $4.84^{+1.10}_{-1.08}$~amu. Key contributor species to this value are CH$_4$ and CO$_2$, which have well-constrained posteriors, as well as H$_2$O and C$_2$H$_6$, whose posteriors display a low abundance tail but still have significant probability density at 1\% and even some above 10\% mixing ratio. Including C$_2$H$_6$ in the model, with a molecular mass of 30.1~amu, therefore has a significant influence on the atmospheric MMW. This is demonstrated by the lower MMW constraint of $4.10^{+1.24}_{-1.15}$~amu obtained when C$_2$H$_6$ is removed from the model, and the excess absorption at $\sim$3.5~$\mu$m is instead explained with DMS at much lower abundance. As such, the MMW constraint depends on which methyl-bearing species is invoked to explain the observations. Lastly, we note that both H$_2$O and C$_2$H$_6$ which significantly influence the MMW constraint, are inferred at a tentative $\sim$2~$\sigma$ level. Therefore, caution is advised when interpreting the MMW constraints of TOI-270~d with current data.

\begin{table*}[]
    \centering
    \addtolength{\tabcolsep}{-0.4em}
    \caption{Constraints on the cloud parameters, reference pressure, NIRISS-NIRSpec relative offset and inferred mean-molecular weight.}
    \vspace{-1mm}
    \def\arraystretch{1.4}
    \begin{tabular}{l||ccccccc}
        Model & $f_\mathrm{c}$ & $\mathrm{log}(a)$ & $\gamma$ & $\mathrm{log}(P_\mathrm{c} / \mathrm{bar}) $ & $\mathrm{log}(P_\mathrm{ref} / \mathrm{bar}) $ & Offset (ppm) & MMW (amu) \\
        \hline
        \hline

        Canonical & $0.82^{+0.12}_{-0.20}$ & $1.11^{+3.44}_{-3.29}$ & $-11.33^{+6.94}_{-3.29}$ & $-2.09^{+0.50}_{-0.48}$ & $-3.41^{+0.25}_{-0.24}$ & $27.8^{+12.4}_{-12.6}$ & $4.84^{+1.10}_{-1.08}$\\
        No C$_2$H$_6$ & $0.86^{+0.09}_{-0.15}$ & $0.87^{+3.48}_{-3.15}$ & $-11.23^{+6.85}_{-5.80}$ & $-2.37^{+0.37}_{-0.43}$ & $-3.48^{+0.24}_{-0.25}$  & $29.4^{+13.1}_{-12.9}$ & $4.10^{+1.24}_{-1.15}$ \\ 

        No C$_2$H$_6$, DMS & $0.87^{+0.09}_{-0.15}$ & $0.89^{+3.47}_{-3.15}$ & $-11.26^{+7.11}_{-5.78}$ & $-2.43^{+0.37}_{-0.43}$ & $-3.51^{+0.24}_{-0.23}$ & $31.7^{+13.4}_{-13.4}$ & $4.55^{+1.04}_{-1.06}$ \\ 
        \hline
        Model & $f_\mathrm{c}$ &$\mathrm{log}(X_\mathrm{aerosol})$ & $\mathrm{log}(r_\mathrm{c} / \mu m)$ & $H_\mathrm{c}$ & $\mathrm{log}(P_\mathrm{ref} / \mathrm{bar}) $ & Offset (ppm) & MMW (amu)  \\ 
        \hline
         Liquid H$_2$O Aerosols & $0.66^{+0.23}_{-0.30}$ & $-12.44^{+4.54}_{-3.63}$ & $-0.85^{+0.47}_{-0.79}$ & $0.51^{+0.25}_{-0.21}$ & $-3.17^{+0.39}_{-0.31}$ & $21.1^{+12.2}_{-11.4}$ & $ 4.59^{+1.80}_{-1.77}$ \\ 
         Ice H$_2$O Aerosols & $0.64^{+0.25}_{-0.31}$ & $-12.40^{+4.57}_{-3.76}$ & $-0.84^{+0.50}_{-0.95}$ & $0.49^{+0.23}_{-0.20}$ & $-3.12^{+0.36}_{-0.32}$ & $21.7^{+10.7}_{-11.6}$ & $5.35^{+1.73}_{-1.40}$ \\
         Titan-like Tholin Aerosols & $0.76^{+0.17}_{-0.26}$ & $-11.25^{+3.40}_{-3.92}$ & $-1.20^{+0.66}_{-0.94}$ & $0.56^{+0.22}_{-0.16}$ & $-3.26^{+0.30}_{-0.26}$ & $23.4^{+12.3}_{-12.0}$ & $ 5.13^{+1.34}_{-1.05}$ \\ 
        \hline
    \end{tabular}
    \begin{flushleft}
    \footnotesize{ \textbf{Note.} The above constraints are obtained with the canonical retrieved atmospheric model, more restricted models with various combinations of methyl-bearing species absent, and models with Mie-scattering aerosols instead of parametric clouds and hazes. $X_\mathrm{Aerosol}$ denotes the mixing ratio of the given aerosol species in each model. Also shown are the inferred MMW constraints, computed from the retrieved mixing ratio posterior distributions for all species present in each model.}
    \end{flushleft}
    \label{tab:additional_constraints}
\end{table*}

\subsection{Unocculted stellar heterogeneities}

We carried out additional retrievals which include the spectral distortion induced by unocculted stellar heterogeneities, as detailed in Section \ref{sec:methods}. The retrieved heterogeneity parameters indicate that the spectrum is not significantly affected by stellar heterogeneities. The retrieved heterogeneity temperature is $3353^{+151}_{-194}$~K, which is slightly lower but consistent with the photospheric temperature of $3504^{+88}_{-85}$~K. Given how close the two temperatures are, distortions from differences between the spectra of the heterogeneities and the pristine photosphere are minor. Moreover, the heterogeneity coverage fraction is piled against the edge of the prior at 0, with a median and 1-$\sigma$ estimate of $0.05^{+0.04}_{-0.02}$.

Beyond the stellar heterogeneities, the retrieved abundance constraints are comparable to those obtained with the canonical retrievals above. CH$_4$ is constrained to a  log-mixing ratio of$-1.81^{+0.28}_{-0.31}$ while CO$_2$ is constrained to $-1.55^{+0.27}_{-0.53}$ and CS$_2$ to $-4.60^{+0.61}_{-0.87}$. C$_2$H$_6$ once again has a posterior distribution with a tail towards low abundances, and is constrained to a log-mixing ratio of $-2.22^{+0.85}_{-1.37}$. The temperature at 10~mbar meanwhile is constrained to $319^{+53}_{-47}$~K.

The inclusion of stellar heterogeneities in the retrieved model does not improve the fit to the data enough to justify the increase in parameter space. Specifically, the canonical model with heterogeneities is only favoured at a $1.1 \pm 0.1 \sigma$ level, indicating no statistically significant preference. Therefore, based on both the retrieved constraints for the heterogeneity parameters and the model evidence, we conclude that the observed transmission spectrum is not significantly impacted by unocculted stellar heterogeneities, consistent with the findings of \citetalias{Holmberg2024}.

\subsection{Water, ice, and tholin aerosols}\label{subsec:mie_aerosols}

The canonical retrievals presented above indicate that atmospheric aerosols may be affecting the observed spectrum to some extent, as evidenced by the $P_\mathrm{c}$ constraints. We now examine what information about the physical properties of these aerosols can be extracted, by replacing the parametric clouds and hazes model with a Mie-scattering calculation consisting of H$_2$O in both ice and liquid states, as well as Titan-like tholins. 

The retrieved molecular abundances and atmospheric temperature for each aerosol are summarised in Table \ref{tab:constraints}, while the aerosol parameter constraints are presented in Table \ref{tab:additional_constraints}. In all three cases, we retrieve similar abundance constraints as the canonical case. We retrieve a slightly lower CH$_4$ abundance constraint than in the canonical case, but still within 1-$\sigma$, while the retrieved constraints for CO$_2$, H$_2$O, CS$_2$, and C$_2$H$_6$ are all similar to their respective canonical constraints and well within 1-$\sigma$. The inferred MMW constraints are all somewhat higher than in the canonical case but also well within 1-$\sigma$, and similarly the retrieved temperatures are all consistent with the canonical case. For all three cases, we find that CH$_3$OH, another methyl-bearing species, also displays a peaked posterior distribution but with a prominent tail towards lower abundances, while DMS is unconstrained.

For example, the retrieval with ice H$_2$O aerosols constrains CH$_4$ to a log-mixing ratio of $-2.21^{+0.34}_{-0.46}$, CO$_2$ to $-1.74^{+0.41}_{-0.79}$, H$_2$O to $-2.21^{+0.34}_{-0.46}$, CS$_2$ to $-4.88^{+0.71}_{-0.94}$, and C$_2$H$_6$ to $-2.22^{+0.92}_{-1.26}$. We retrieve a 10~mbar temperature of $295^{+60}_{-58}$~K while the overall inferred MMW is constrained to  $5.13^{+1.34}_{-1.05}$~amu.

For all three cases, we obtain peaked but unconstrained posteriors for their physical properties, presented in Table \ref{tab:additional_constraints}. Again considering the ice H$_2$O retrieval as an example, we find an aerosol mixing ratio constraint of $-12.40^{+4.57}_{-3.76}$, a modal particle size constraint of $\mathrm{log(}r_c /\mu\mathrm{m)} = -0.84^{+0.50}_{-0.95}$, vertical scale height of $0.49^{+0.23}_{-0.20}$ and terminator coverage fraction of $0.64^{+0.25}_{-0.31}$.

As such, we find that while there are nominal indications that atmospheric aerosols are impacting the observed transmission spectrum, as evidenced by the peaked posterior distributions obtained for their physical properties, as well as the constraints for the parametric clouds and hazes model above, their spectral contributions are not substantial enough to result in definitive constraints for their physical properties.

\subsection{Comparison with B24 and HM24}

Relative to \citetalias{Benneke2024} the present analysis results in a lower atmospheric temperature as well as lower atmospheric abundances for all species and consequently a lower MMW constraint. Specifically, the constraints for CH$_4$, CO$_2$, and H$_2$O are lower but within 1-$\sigma$ of the constraints obtained by \citetalias{Benneke2024}. The CS$_2$ constraint meanwhile is more that 1-$\sigma$ lower than that of \citetalias{Benneke2024}. The overall atmospheric MMW of $4.84^{+1.10}_{-1.08}$~amu obtained with our canonical retrieval is lower than the $5.47^{+1.25}_{-1.14}$~amu obtained by \citetalias{Benneke2024} but consistent to within 1-$\sigma$. However, the retrieved MMW depends on the methyl species considered in the retrieval. For example, the canonical model with C$_2$H$_6$ removed, leads to an MMW of $4.10^{+1.24}_{-1.15}$. Lastly, the 10~mbar temperature of $323^{+58}_{-52}$~K obtained with our canonical retrieval is significantly lower than the $385^{+44}_{-42}$~K value obtained by \citetalias{Benneke2024} but has overlapping 1-$\sigma$ uncertainties.

At the same time, we revise the constraints obtained by \citetalias{Holmberg2024} without the NIRISS observations. \citetalias{Holmberg2024} consider the present NIRSpec data along with HST/WFC3 data in the 1.1-1.7$\mu$m range. The broader wavelength coverage of the current data result in more precise abundance constraints. Our CO$_2$ and H$_2$O constraints are consistent with \citetalias{Holmberg2024}, while the CH$_4$ constraint of $-1.86^{+0.30}_{-0.29}$ obtained in this work is over 1-$\sigma$ higher than the $-2.72^{+0.41}_{-0.50}$ of \citetalias{Holmberg2024}. Conversely, the present CS$_2$ constraint of $-4.47^{+0.65}_{-1.10}$ is more than 1-$\sigma$ lower than the value of $-3.07^{+0.74}_{-0.91}$ from \citetalias{Holmberg2024}. Meanwhile the 10~mbar temperature retrieved in this work is slightly higher but within 1-$\sigma$ of the 10~mbar constraint obtained by \citetalias{Holmberg2024} of $298^{+80}_{-75}$~K. Our canonical MMW constraint is consistent with the $\sim$2.4-4.8~amu range estimated from the abundance constraints of \citetalias{Holmberg2024}.

\section{Summary and conclusion}\label{sec:discussion}

In this work we conducted a comprehensive atmospheric characterisation of TOI-270d to address three goals: (i) resolve the debate over its atmospheric properties, (ii) present revised atmospheric constraints using all available JWST observations, and (iii) explore additional constraints on the atmospheric properties. We achieve this by combining JWST NIRISS and NIRSpec transmission spectroscopy over the 1-5 $\mu$m range. 

We first assessed the possibility of a mixed envelope scenario as inferred by \citetalias{Benneke2024}, which involves a relatively high terminator temperature and an atmosphere with high abundances of molecular species like H$_2$O and CH$_4$. We find in Section \ref{sec:benneke_reproduction} that the retrieved atmospheric constraints of \citetalias{Benneke2024} were affected primarily by the omission of methyl species, which are required to explain a part of the observations. Using the data presented by \citetalias{Benneke2024} and an equivalent atmospheric model with the additional inclusion of methyl species and the reference pressure as free parameters, we retrieve an isothermal terminator temperature of $251^{+52}_{-59}$~K and an inferred MMW of $3.79^{+1.37 }_{-1.10}$ amu, which are both lower than the $385^{+44}_{-42}$~K temperature and $5.47^{+1.25}_{-1.14}$~amu MMW obtained by \citetalias{Benneke2024}.

We subsequently applied our canonical retrieval framework to NIRSpec observations presented by \citetalias{Holmberg2024} combined with a new reduction of NIRISS observations. Similarly to \citetalias{Benneke2024} and \citetalias{Holmberg2024}, we obtain confident detections of CH$_4$ at a 6.4$\sigma$ level and CO$_2$ at $3.9 \sigma$, as well as a tentative inference of H$_2$O at $2.1 \sigma$, and CS$_2$ at $2.0 \sigma$. We present revised constraints for the atmospheric composition of TOI-270~d, with log-mixing ratios of $-1.86^{+0.30}_{-0.29}$ for CH$_4$, $-1.71^{+0.38}_{-0.66}$ for CO$_2$, $-1.88^{+0.78}_{-4.13}$ for H$_2$O and $-4.74^{+0.65}_{-1.10}$ for CS$_2$. We also constrain the atmospheric temperature at 10~mbar to $323^{+58}_{-52}$~K. 

Lastly, we consider what additional atmospheric constraints are possible with the present data. We find evidence for methyl-bearing species being present in the atmosphere of TOI-270~d at a $2.1-2.5 \sigma$ level. In the present study, we consider C$_2$H$_6$, DMS, and several other species. We find that C$_2$H$_6$ and, in its absence, DMS can give rise to the observed excess atmospheric absorption at $\sim$3.5~$\mu$m. We note that while there is tentative evidence for the presence of methyl-bearing species, the present data is not sufficient to uniquely determine which methyl-bearing species or group of species is giving rise to the observed absorption features.

Our present atmospheric constraints lie between those reported by \citetalias{Benneke2024} and \citetalias{Holmberg2024}. Specifically, we retrieve lower abundance constraints for CH$_4$, CO$_2$, H$_2$O, and CS$_2$ than \citetalias{Benneke2024}, resulting in a MMW constraint that is lower but consistent within 1-$\sigma$. The retrieved temperature is also lower than that presented by \citetalias{Benneke2024} but similarly the median values lie within 1-$\sigma$. Compared to \citetalias{Holmberg2024}, who used the present NIRSpec data along with HST WFC3 data, the present analysis results in a CH$_4$ abundance that is higher by more than 1-$\sigma$, consistent CO$_2$ and H$_2$O abundances and a lower CS$_2$ constraint. The 10~mbar temperature retrieved with our canonical model meanwhile is marginally higher but well within 1-$\sigma$ of that presented by \citetalias{Holmberg2024}. The corresponding MMW constraint is also consistent with that estimated from the abundances reported by \citetalias{Holmberg2024}.

\subsection{Chemical tracers of internal structure}\label{subsec:interior}

Several competing hypotheses have been put forward for the internal structure of TOI-270~d and other sub-Neptunes like it. The detection of atmospheric signatures with HST and JWST preclude any scenarios with atmospheres not containing H$_2$ in significant quantities, for instance, super-Earths or water worlds. There are however several hypotheses consistent with H$_2$-rich atmospheres, ranging from mini-Neptunes with deep atmospheres, gas dwarfs with rocky interiors, Hycean worlds with shallower ocean surfaces and miscible-envelope sub-Neptunes.

The present constraints are consistent with the hypothesis of TOI-270~d being a Hycean world \citep{Madhusudhan2021, Evans2023}. As noted above, we find confident detections of CH$_4$ and CO$_2$, as well as placing a low upper limit on the abundance of NH$_3$. This combination is suggestive of a Hycean world, as previously noted by \citetalias{Holmberg2024}. This is also similar to K2-18~b, another temperate sub-Neptune with atmospheric CH$_4$ and CO$_2$ detections and a lack of NH$_3$ \citep{Madhusudhan2023b}. However unlike K2-18~b we do not find strong constraints on the CO abundance, which precludes a robust assessment of the Hycean scenario.

An alternative hypothesis for TOI-270~d has been put forward by \citetalias{Benneke2024}, who classify the planet as a miscible-envelope sub-Neptune, that is, one with no surface and a well-mixed, high-MMW atmosphere. This conclusion hinges on a relatively high retrieved terminator temperature and high inferred atmospheric MMW. As shown in Section \ref{sec:benneke_reproduction}, these retrieved properties are affected by the absence of methyl-group species in the atmospheric model of \citetalias{Benneke2024}. Secondly, we find a lower H$_2$O abundance and terminator temperature than \citetalias{Benneke2024}, which along with the non-detection of NH$_3$ makes it difficult to reconcile the mini-Neptune scenario.

As noted previously, the present constraints on CH$_4$, CO$_2$, and NH$_3$ are comparable to those obtained for K2-18~b, another temperate sub-Neptune and candidate Hycean world \citep{Madhusudhan2023b}. It has been previously shown that the Hycean nature of K2-18~b hinges on the specific Bond albedo of its dayside, which needs to be high enough to avoid a runaway greenhouse effect \citep{Scheucher2020, Piette2020, innes2023runaway, Pierrehumbert2023}. TOI-270~d has a higher instellation than K2-18~b, leading to an equilibrium temperature that is higher by over $\sim$50~K, depending on the Bond albedo values assumed for the two planets. We note however that this higher temperature does not preclude TOI-270~d being a Hycean world, as such conditions may still prevail globally for sufficiently high Bond albedo values, or on its nightside or elsewhere for sufficiently low redistribution efficiencies. In the latter case, TOI-270~d would be a Dark Hycean world \citep{Madhusudhan2021}. The retrieved terminator temperature in this work is consistent with such scenarios.

\subsection{Presence of methyl-bearing species}

As described in Section \ref{subsec:detections}, we find evidence for the presence of methyl-bearing species C$_2$H$_6$ and/or DMS at similar or greater confidence than CS$_2$ and H$_2$O. This evidence persists even when we consider alternative reductions of the same observations, as shown in Section \ref{subsec:bennekedata_modelcomparison}. In principle, other methyl-bearing species not considered in this work may also be responsible wholly or in part for the observed excess absorption. Despite this degeneracy, the present retrievals provide only tentative estimates of their abundances, if they are indeed present, with log-mixing ratio constraints of $-2.20^{+0.87}_{-2.01}$ for C$_2$H$_6$ and $-5.80^{+1.00}_{-3.71}$ for DMS instead of C$_2$H$_6$.

A mixing ratio of 1\% would mean that C$_2$H$_6$ is present at a comparable abundance to CH$_4$, the molecule expected to be the primary C-carrier for such atmospheres. This would imply a highly efficient pathway for the production of C$_2$H$_6$ beyond equilibrium chemistry, or conversely, highly ineffective pathways for the removal of C$_2$H$_6$ from the atmosphere. Tentative indications of DMS have been found in the atmosphere of K2-18~b \citep{Madhusudhan2023b}, at mixing ratios consistent with those indicated in this work. DMS has been identified as a secondary biomarker on Earth, produced by phytoplankton. Assessing the physical plausibility of DMS being present necessitates further exploration of possible abiogenic sources for the exotic conditions prevailing in the atmospheres of temperate sub-Neptune planets like TOI-270~d.

\subsection{Future outlook}

 Future observations of TOI-270~d can help improve the present constraints on its atmospheric properties, enabling more confident detections of molecular species and more precise abundance estimates. Moreover, with more precise data obtained by co-adding additional transits, it may be possible to place more stringent constraints on the planet's clouds and hazes properties.

In particular, more observations can potentially be effective in resolving the present degeneracy between methyl-bearing species seen in the present work. At the same time, it is important to expand the library of absorption linelists for such molecules, especially in the H$_2$-dominated regime found in temperate sub-Neptune exoplanets. Such species can serve as informative tracers for the planet's internal structure and surface characteristics, as described above. Altogether, the present and upcoming observations are set to cement the status of TOI-270~d as an important benchmark in the exploration of the temperate sub-Neptune regime.

\vspace{2mm}
\begin{acknowledgements}
This work is based on observations made with the NASA/ESA/CSA James Webb Space Telescope as part of Cycle 2 GO Program 4098 (PI: B. Benneke). This work is supported by research grants to N.M. from the UK Research and Innovation (UKRI) Frontier Grant (EP/X025179/1), the MERAC Foundation, Switzerland, and the UK Science and Technology Facilities Council (STFC) Center for Doctoral Training (CDT) in Data Intensive Science at the University of Cambridge (STFC grant No. ST/P006787/1). N.M. and M.H. acknowledge support from STFC and the MERAC Foundation toward the doctoral studies of M.H.

This work was performed using resources provided by the Cambridge Service for Data Driven Discovery operated by the University of Cambridge Research Computing Service (\url{www.csd3.cam.ac.uk}), provided by Dell EMC and Intel using Tier-2 funding from the Engineering and Physical Sciences Research Council (capital grant EP/P020259/1), and DiRAC funding from STFC (\url{www.dirac.ac.uk}).\\[4pt]

{\it Author Contributions. } N.M. conceived and planned the project. S.C. led the execution of the project with guidance from N.M., conducted the atmospheric retrieval analysis and led the writing of the manuscript. M.H. conducted the NIRISS data reduction and light-curve analysis and contributed the corresponding text. 

\end{acknowledgements}

\bibliography{refs}{}
\bibliographystyle{aa.bst}

\begin{appendix}

\section{Canonical atmospheric model constraints}

\begin{figure*}
\noindent

\includegraphics[angle=0,width=\textwidth]{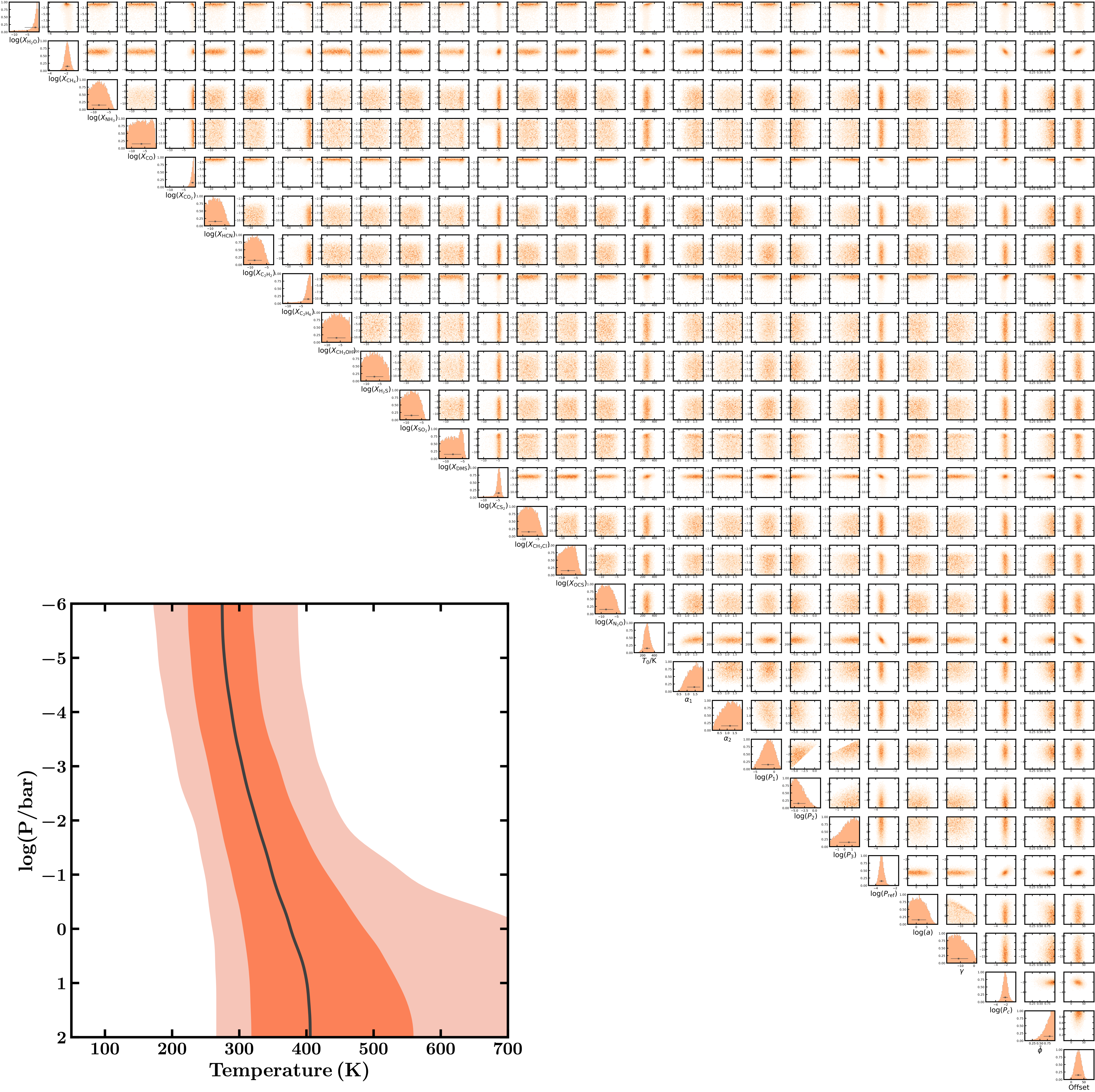}
\caption{Posterior probability distribution obtained from the canonical retrieval described in Sections \ref{sec:methods} and \ref{sec:canonical_retrievals} and corresponding pressure-temperature profile. The circles with horizontal error bars in the corner plot denote the median and corresponding 1-$\sigma$ intervals of each 1D distribution. The black curve denotes the median pressure-temperature profile, inferred from the temperature profile probability distributions. The darker and lighter orange shaded regions denote the corresponding 1- and 2-$\sigma$ intervals, respectively.}
\label{fig:canonical_posteriors_full}
\end{figure*}

Figure \ref{fig:canonical_posteriors_full} shows the complete posterior probability distribution for our canonical model described in Section \ref{sec:methods}. Also shown is the corresponding retrieved temperature profile, which is consistent with an isotherm to within 1-$\sigma$. A subset of the individual marginalised 1D posterior distributions are also shown as orange posterior distributions in Figure \ref{fig:canonical_posteriors}.

Table \ref{tab:retrieval_priors} shows all free parameters considered in the canonical atmospheric model and the model excluding C$_2$H$_6$. Also shown are the corresponding Bayesian prior distributions and posterior constraints. For each parameter we show the constraint derived using the median and percentile-based approach as used throughout this work. We also show the parameter estimates derived using the mode and the highest posterior density interval (HPDI) approach, corresponding to the smallest interval enclosing a given probability, for example, 68.3\% for 1-$\sigma$ intervals.

\begin{table*}
\small
\def\arraystretch{1.4}
\caption{Retrieved atmospheric parameters and prior probability distributions.}
\vspace{-2mm}
\begin{tabular}{lclcccc} \hline \hline
    Parameter & Bayesian Prior  & Description & \multicolumn{2}{c}{Canonical Model} & \multicolumn{2}{c}{Without C$_2$H$_6$}\\[0.5mm]
    \hline
    & & & Percentile & HPDI & Percentile & HPDI \\[0.5mm]
    \hline
    $\mathrm{log}(X_\mathrm{H_2O})$ &  $\mathcal{U}$(-12, -0.3)  &  Mixing ratio of H$_2$O           &  $-1.88_{-4.13}^{+0.78}$  & $-1.23^{+0.55}_{-1.68}$ &  $-1.52_{-4.07}^{+0.53}$  &  $-1.00^{+0.31}_{-1.34}$\\[0.5mm]
    $\mathrm{log}(X_\mathrm{CH_4})$   &  $\mathcal{U}$(-12, -0.3)  &  Mixing ratio of CH$_4$         &  $-1.86_{-0.29}^{+0.30}$  & $-1.82^{+0.28}_{-0.32}$ &  $-1.77_{-0.32}^{+0.30}$  &  $-1.85^{+0.39}_{-0.23}$\\[0.5mm]
    $\mathrm{log}(X_\mathrm{NH_3})$   &  $\mathcal{U}$(-12, -0.3)  &  Mixing ratio of NH$_3$         &  $-8.22_{-2.40}^{+2.39} (-4.24)$  & $-6.61^{+0.65}_{-4.14}$ &  $-8.14_{-2.45}^{+2.53} (-4.02)$ &  $-7.47^{+1.74}_{-3.22}$\\[0.5mm]
    $\mathrm{log}(X_\mathrm{CO})$  &  $\mathcal{U}$(-12, -0.3)  &  Mixing ratio of CO                &  $-6.30_{-3.64}^{+3.59} (-1.32)$  & $-2.91^{+1.44}_{-5.64}$ &  $-6.36_{-3.77}^{+3.81} (-1.28)$  &  $-2.29^{+1.01}_{-6.36}$\\[0.5mm]
    $\mathrm{log}(X_\mathrm{CO_2})$ &  $\mathcal{U}$(-12, -0.3)  &  Mixing ratio of CO$_2$           &  $-1.71_{-0.66}^{+0.38}$  & $-1.50^{+0.32}_{-0.57}$ &  $-1.89_{-0.77}^{+0.50}$  &  $-1.36^{+0.19}_{-0.94}$\\[0.5mm]
    $\mathrm{log}(X_\mathrm{HCN})$   &  $\mathcal{U}$(-12, -0.3) &  Mixing ratio of HCN              &  $-8.33_{-2.35}^{+2.43} (-4.28)$ & $-9.36^{+3.38}_{-1.39}$ &  $-8.19_{-2.46}^{+2.52} (-4.08)$  &  $-8.37^{+2.46}_{-2.48}$\\[0.5mm]
    $\mathrm{log}(X_\mathrm{C_2H_2})$   &  $\mathcal{U}$(-12, -0.3)  &  Mixing ratio of C$_2$H$_2$   &  $-8.64_{-2.16}^{+2.18} (-5.01)$  & $-8.16^{+1.74}_{-2.59}$ &  $-8.51_{-2.22}^{+2.20} (-4.90)$  &  $-9.82^{+3.38}_{-1.03}$\\[0.5mm]
    $\mathrm{log}(X_\mathrm{C_2H_6})$   &  $\mathcal{U}$(-12, -0.3)  &  Mixing ratio of C$_2$H$_6$   &  $-2.20_{-2.01}^{+0.87}$  & $-1.81^{+0.88}_{-1.12}$ &  -   &  -\\[0.5mm]
        $\mathrm{log}(X_\mathrm{CH_3OH})$   &  $\mathcal{U}$(-12, -0.3)  &  Mixing ratio of CH$_3$OH  &  $-6.94_{-3.51}^{+3.40} (-1.46)$  & $-4.53^{+1.8}_{-5.07}$ &  $-6.65_{-3.47}^{+3.44} (-1.56)$  &  $-4.99^{+1.74}_{-5.15}$\\[0.5mm]
    $\mathrm{log}(X_\mathrm{H_2S})$   &  $\mathcal{U}$(-12, -0.3)  &  Mixing ratio of H$_2$S         &  $-6.94_{-3.18}^{+3.28} (-1.80)$   & $-7.59^{+3.49}_{-2.93}$ &  $-6.73_{-3.30}^{+3.43} (-1.66)$  &  $-8.64^{+5.37}_{-1.35}$\\[0.5mm]
    $\mathrm{log}(X_\mathrm{SO_2})$   &  $\mathcal{U}$(-12, -0.3)  &  Mixing ratio of SO$_2$         &  $-8.25_{-2.39}^{+2.36} (-4.46)$  & $-7.20^{+1.23}_{-3.51}$ &  $-8.20_{-2.43}^{+2.42} (-4.35)$   &  $-7.92^{+1.70}_{-3.12}$\\[0.5mm]
    $\mathrm{log}(X_\mathrm{DMS})$   &  $\mathcal{U}$(-12, -0.3)  &  Mixing ratio of DMS             &  $-7.84_{-2.67}^{+2.43} (-4.54)$  & $-5.12^{+0.43}_{-4.42}$ &  $-5.80_{-3.71}^{+1.00} (-4.30)$   &  $-5.04^{+0.75}_{-2.75}$\\[0.5mm]
    $\mathrm{log}(X_\mathrm{CS_2})$ &  $\mathcal{U}$(-12, -0.3)  &  Mixing ratio of CS$_2$           &  $-4.74_{-1.10}^{+0.65} (-3.48)$  & $-4.63^{+0.79}_{-0.73}$ &  $-4.71_{-1.14}^{+0.65} (-3.52)$   &  $-4.44^{+0.57}_{-0.97}$\\[0.5mm]
    $\mathrm{log}(X_\mathrm{CH_3Cl})$   &  $\mathcal{U}$(-12, -0.3) &  Mixing ratio of CH$_3$Cl      &  $-7.99_{-2.55}^{+2.60} (-3.85)$  & $-7.14^{+1.44}_{-3.66}$ &  $-7.56_{-2.83}^{+2.85} (-3.41)$   &  $-7.19^{+3.03}_{-2.60}$\\[0.5mm]
    $\mathrm{log}(X_\mathrm{OCS})$  &  $\mathcal{U}$(-12, -0.3) &  Mixing ratio of OCS               &  $-7.65_{-2.71}^{+2.42} (-3.86)$  & $-7.71^{+3.01}_{-2.01}$ &  $-7.60_{-2.83}^{+2.85} (-3.72)$   &  $-5.09^{+0.28}_{-4.90}$\\[0.5mm]
    $\mathrm{log}(X_\mathrm{N_2O})$  &  $\mathcal{U}$(-12, -0.3)  &  Mixing ratio of N$_2$O          &  $-8.45_{-2.24}^{+2.35} (-4.59)$  & $-9.82^{+3.47}_{-1.08}$ &  $-8.34_{-2.34}^{+2.41} (-4.37)$   &  $-9.26^{+3.02}_{-1.69}$\\[0.5mm]
    $T_0 / \mathrm{K} $  &  $\mathcal{U}$(50, 600) &  Temperature at 1 $\mu$bar                       &  $273_{-48}^{+51}$  & $272^{+45}_{-54}$ &  $285_{-47}^{+51}$   &  $285^{+40}_{-56}$\\[0.5mm]
    $T_{10 \mathrm{mbar}} / \mathrm{K}$  &  -  &  Temperature at 10 mbar                             &  $323^{+58}_{-52}$  & $310^{+58}_{-50}$ &  $335^{+58}_{-49}$  &  $325^{+60}_{-48}$\\[0.5mm]
    $\alpha_1 / \mathrm{K}^{-\frac{1}{2}}$   &  $\mathcal{U}$(0.02, 2.00)  &  $P$-$T$ profile curvature  &  $1.44_{-0.44}^{+0.37}$  & $1.66^{+0.28}_{-0.50}$ &  $1.45_{-0.44}^{+0.36}$  &  $1.68^{+0.32}_{-0.44}$\\[0.5mm]
    $\alpha_2/ \mathrm{K}^{-\frac{1}{2}}$ &  $\mathcal{U}$(0.02, 2.00)   &  $P$-$T$ profile curvature  &  $1.19_{-0.53}^{+0.59}$  & $1.02^{+0.85}_{-0.22}$ &  $1.17_{-0.61}^{+0.54}$   &  $1.13^{+0.74}_{-0.37}$\\[0.5mm]
    $\mathrm{log}(P_1/\mathrm{bar})$    &  $\mathcal{U}$(-6, 2)  &  $P$-$T$ profile region limit     &  $-1.59_{-1.74}^{+1.60}$  & $-1.59^{+1.79}_{-1.53}$ &  $-1.61_{-1.69}^{+1.59}$  &  $-2.06^{+2.16}_{-1.11}$\\[0.5mm]
    $\mathrm{log}(P_2/\mathrm{bar})$   &  $\mathcal{U}$(-6, 2)  &  $P$-$T$ profile region limit      &  $-4.04_{-1.29}^{+1.77}$  & $-4.36^{+1.17}_{-1.61}$ &  $-4.04_{-1.31}^{+1.73}$  &  $-4.43^{+1.24}_{-1.53}$\\[0.5mm]
    $\mathrm{log}(P_3/\mathrm{bar})$    &  $\mathcal{U}$(-2, 2) &  $P$-$T$ profile region limit      &  $-0.57_{-1.32}^{+0.97}$  & $1.26^{+0.73}_{-1.28}$ &  $-0.58_{-1.34}^{+0.96}$  &  $1.46^{+0.54}_{-1.51}$\\[0.5mm]
    $\mathrm{log}(P_\mathrm{ref}/\mathrm{bar})$    &  $\mathcal{U}$(-6, 2)  &  Reference pressure at R$_\mathrm{P}$  &  $-3.41_{-0.24}^{+0.25}$  & $-3.39^{+0.18}_{-0.31}$ &  $-3.48_{-0.25}^{+0.24}$  &  $-3.53^{+0.30}_{-0.19}$\\[0.5mm]
    $\mathrm{log}(a)$   &  $\mathcal{U}$(-4, 10) &  Rayleigh enhancement                      &  $1.11_{-3.29}^{+3.44}$  & $1.30^{+2.36}_{-4.29}$ &  $0.87_{-3.15}^{+3.48}$  &  $-0.71^{+4.09}_{-2.31}$\\[0.5mm]
    $\gamma$    &  $\mathcal{U}$(-20, 2) &  Scattering slope                                         &  $-11.33_{-5.72}^{+6.94}$  & $-15.91^{+8.24}_{-3.67}$ &  $-11.23_{-5.80}^{+6.85}$  &  $-16.07^{+9.25}_{-2.8}$\\[0.5mm]
    $\mathrm{log}(P_\mathrm{c}/\mathrm{bar})$   &  $\mathcal{U}$(-6, 2) &  Cloud top pressure        &  $-2.09_{-0.48}^{+0.50}$  & $-2.25^{+0.62}_{-0.36}$ &  $-2.37_{-0.43}^{+0.37}$  &  $-2.35^{+0.38}_{-0.41}$\\[0.5mm]
    $f_\mathrm{c}$   &  $\mathcal{U}$(0, 1) &  Cloud and haze fraction                                  &  $0.82_{-0.20}^{+0.12}$  & $0.95^{+0.05}_{-0.21}$ &  $0.86_{-0.15}^{+0.09}$  &   $0.96^{+0.04}_{-0.16}$\\[0.5mm]
    $\delta_\mathrm{NIRSpec} / \mathrm{ppm}$   &  $\mathcal{U}$(-250, 250) &  NIRISS dataset offset  &  $28_{-13}^{+12}$   & $33^{+8}_{-17}$ &  $29_{-13}^{+13}$  &  $32^{+11}_{-15}$\\[0.5mm]

    \hline
\end{tabular}
\vspace{2mm}
\newline
\footnotesize{\textbf{Note.} For the parameter estimates, the central values and uncertainties in the Percentile columns correspond to the median and 16-84\% percentiles (1-$\sigma$), with values in brackets corresponding to the 2-$\sigma$ upper limits. The equivalent values in the Highest Posterior Density Interval (HPDI) columns correspond to the mode and HPDI for 68.27\% of the probability. The temperature at 10~mbar, which corresponds to the observed photosphere, is derived from the retrieved $P-T$ profile. }
\label{tab:retrieval_priors}
\end{table*}

\section{Additional robustness checks}
\label{sec:robustness_checks}
We present an additional retrieval using UltraNest \citet{Buchner2021}, an alternative Bayesian nested sampling implementation, to assess the robustness of the results we present in the main body of the present work, both in terms of the retrieved posterior distributions and computed model evidence. It has been previously reported that UltraNest provides more accurate model evidence values than MultiNest, albeit at higher computational cost \citep[e.g.][]{Buchner2016, Buchner2021}.

\begin{figure*}
\noindent
\includegraphics[angle=0,width=\textwidth]{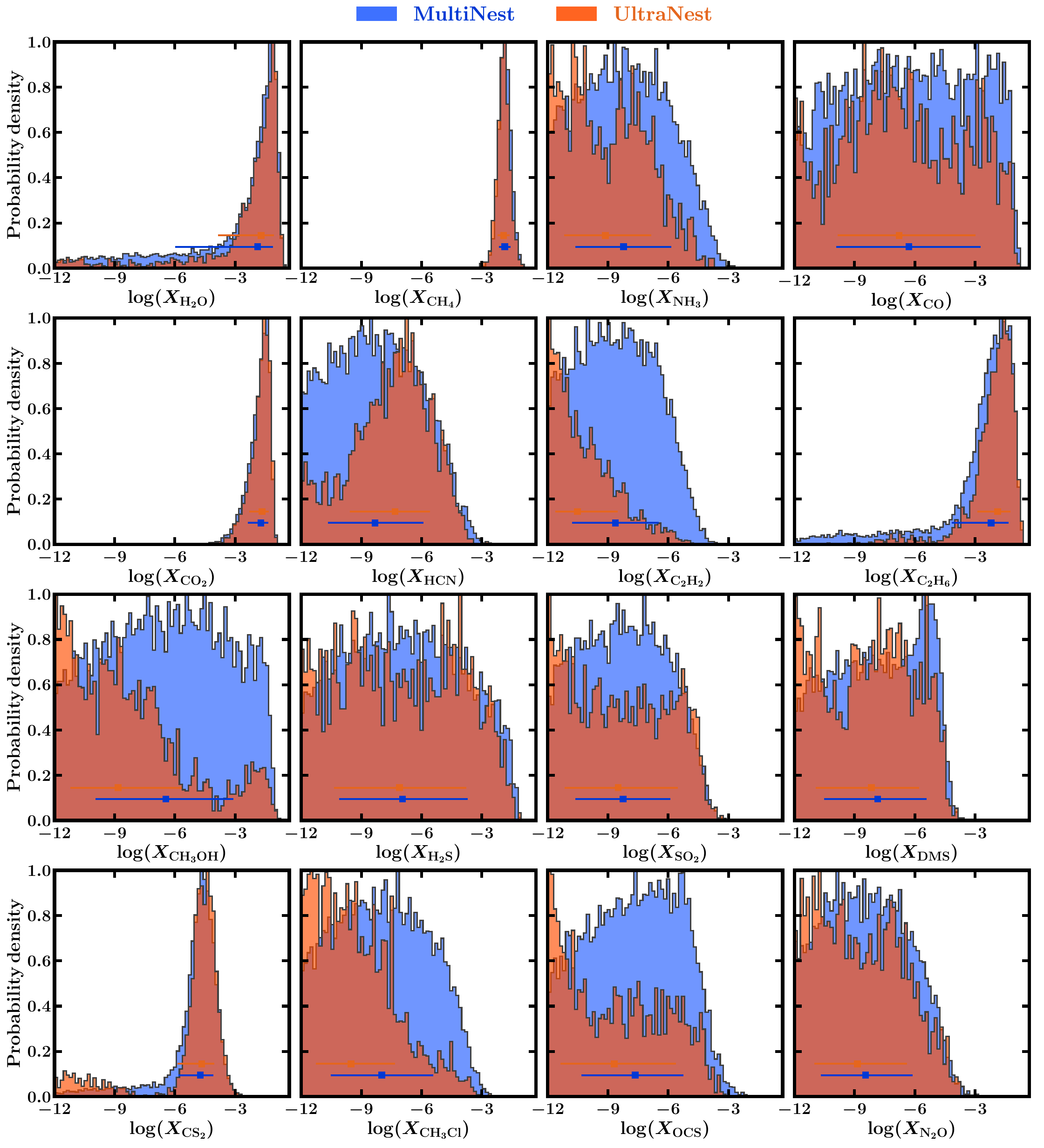}
\caption{ Mixing ratio posterior distributions obtained using the canonical atmospheric model described in Section \ref{sec:methods} using the MultiNest (orange) and UltraNest (blue) nested sampling implementations. The orange posterior distributions for the MultiNest retrieval are the same as the ones shown in Figures \ref{fig:canonical_posteriors} and \ref{fig:canonical_posteriors_full} Horizontal orange and blue points with errorbars denote the median and 1-$\sigma$ intervals for the posterior distribution of the same colour.}
\label{fig:NS_comparison}
\end{figure*}

We adapted the canonical retrieval presented in Section \ref{sec:methods}, with the only difference being the replacement of the nested sampling algorithm. The retrieved posterior distributions for a number of chemical species is shown in Figure \ref{fig:NS_comparison}. The two nested sampling implementations obtain consistent median and 1-$\sigma$ values, especially for species with well-constrained posteriors. Notable differences are the decreased probability towards the center of the prior space for some species such as H$_2$S and SO$_2$, which are otherwise unconstrained for both retrievals. As such, in the case of well-constrained abundances, MultiNest produces constraints that are consistent with UltraNest, but in the case of unconstrained or weakly-constrained species, MultiNest may potentially be producing inaccurate posterior distributions.

We carried out additional retrievals with UltraNest where molecules are excluded from the model to calculate the Bayesian model preference. We find the model with CH$_4$ is preferred at a $6.19 \pm 0.18 \sigma$ level, which is consistent with the value computed with our canonical retrievals using MultiNest. We therefore conclude that the present results were not significantly affected by any potential biases in the specific implementation of nested sampling in MultiNest. On jointly excluding the tentatively detected methyl-bearing species C$_2$H$_6$ and DMS, we find a model preference for their inclusion at a $2.1 \pm 0.3 \sigma$ level, which is also consistent with the MultiNest result.

\end{appendix}

\end{document}